\documentclass[11pt,a4paper]{article}
\usepackage{jheppub}
\pdfoutput=1

\usepackage{siunitx}
\usepackage{pdfpages}
\usepackage{float}
\usepackage{textcomp}               
\usepackage{braket,slashed,bm}
\usepackage{array,multirow}
\usepackage[normalem]{ulem}
\usepackage[usenames,dvipsnames]{xcolor}
\usepackage{cancel,youngtab}

\usepackage[T1]{fontenc} 
\usepackage{mathrsfs}
\usepackage{booktabs}
\usepackage{adjustbox}
\usepackage{mathtools}
\usepackage{hhline}
\usepackage{soul}

\usepackage{tikz}
\usetikzlibrary{arrows,decorations.pathmorphing,backgrounds,positioning,fit,petri,automata,shadows,calendar,mindmap,
decorations.markings,calc}

\definecolor{labelkey}{rgb}{0,0.5,0.0}

\usepackage{xspace}
\usepackage{slashed}
\usepackage{array,multirow}
\usepackage{graphicx}
\usepackage{graphics}
\usepackage{epsfig}
\usepackage{amsfonts}
\usepackage{amsmath}
\usepackage{amssymb}
\usepackage{dsfont}
\usepackage{color}
\usepackage{pdflscape}
\usepackage{pifont}
\usepackage{pgfplots}
\usepackage{tikz} 
\usepackage{pgfplotstable}
\usepackage{bbold}
\usepackage[capitalise, english]{cleveref}

\newcommand{\lam}{\lambda}

\newcommand{\kap}{\kappa}

\usepackage{bm}                                  

\newcommand{\beq}{\begin{equation}}
\newcommand{\eeq}{\end{equation}}
\newcommand{\be}{\begin{equation}}
\newcommand{\ee}{\end{equation}}
\newcommand{\bea}{\begin{eqnarray}}
\newcommand{\eea}{\end{eqnarray}}
\newcommand{\ben}{\begin{eqnarray*}}
\newcommand{\een}{\end{eqnarray*}}

\newcommand{\bma}{\begin{pmatrix}}
\newcommand{\ema}{\end{pmatrix}}

\def\lixo#1{}

\def\slashchar#1{\setbox0=\hbox{$#1$}           
  \dimen0=\wd0                                    
  \setbox1=\hbox{/} \dimen1=\wd1                  
  \ifdim\dimen0>\dimen1                           
    \rlap{\hbox to \dimen0{\hfil/\hfil}}            
    #1                                             
  \else                                          
    \rlap{\hbox to \dimen1{\hfil$#1$\hfil}}        
    /                                           
 \fi}                                           %

\newcommand\lsim{\lesssim}

\newcommand{\dslash}[1]{#1 \llap{/\kern-0.5pt}}
\newcommand{\Dslash}[1]{#1 \llap{/\kern+1.5pt}}
\newcommand{\DDslash}[1]{#1 \llap{/\kern+2.3pt}}
\newcommand{\dslashh}[1]{#1 \llap{/\kern+1pt}}

\definecolor{ForestGreen}{rgb}{0.0, 0.27, 0.13}

	\preprint{\begin{flushright} BONN-TH-2022-15
	\end{flushright}}	

	\title{Searching for a Single Photon from Lightest Neutralino Decays in R-parity-violating Supersymmetry at FASER
		}

		\author[a]{Herbi~K.~Dreiner,}
		\emailAdd{dreiner@uni-bonn.de}
		\affiliation[a]{Bethe Center for Theoretical Physics \& Physikalisches Institut der Universit\"at Bonn,\\ Nu{\ss}allee 12, 53115 Bonn, Germany}

		\author[a]{Dominik K\"ohler,}
		\emailAdd{koehler@physik.uni-bonn.de}

		\author[a]{Saurabh Nangia,}
		\emailAdd{nangia@physik.uni-bonn.de}

		\author[b,c]{and Zeren~Simon~Wang}
		\emailAdd{wzs@mx.nthu.edu.tw}
		\affiliation[b]{Department of Physics, National Tsing Hua University, Hsinchu 300, Taiwan}
		\affiliation[c]{Center for Theory and Computation, National Tsing Hua University, Hsinchu 300, Taiwan}

\abstract{In this work, we propose a search for a single  photon at \texttt{FASER} and \texttt{FASER2}, produced from decays of bino-like, sub-GeV lightest neutralinos in the theoretical 
framework of the R-parity-violating (RPV) Minimal Supersymmetric Standard Model (MSSM). We consider a list of representative benchmark scenarios with one or two non-vanishing RPV couplings. The photon has an energy $\mathcal{O}\left(0.1\right)-\mathcal{O}\left(1\right)$ TeV. We find a sensitivity reach for RPV couplings beyond the current bounds by orders of magnitude at \texttt{FASER} and \texttt{FASER2}.}

\pgfplotsset{compat=1.8}
\begin{document}
\maketitle

\section{Introduction}\label{sec:introduction}

The discovery of a Higgs boson~\cite{ATLAS:2012yve,CMS:2012qbp} at the Large Hadron Collider 
(LHC) at CERN, Switzerland, has completed the spectrum of the Standard Model (SM) of particle 
physics. Despite the huge successes, the SM provides an incomplete description of the
Universe. For instance, the observed neutrino 
oscillations~\cite{deSalas:2020pgw,Esteban:2020cvm,Capozzi:2017ipn} require massive neutrinos, 
in disagreement with the SM. The fine-tuning problem of the Higgs boson -- or hierarchy
problem~\cite{Gildener:1976ai,Veltman:1980mj} -- is only resolved beyond the SM 
(BSM), \textit{e.g.}, by supersymmetry (SUSY)~\cite{Nilles:1983ge,Martin:1997ns}. Furthermore, dark matter and dark energy, as well as baryogenesis in the early Universe are
all unexplained within the SM. 

Searches for BSM physics have been performed since even prior to the Higgs-boson discovery, on 
various experimental and observational fronts. These probes include colliders, beam-dump experiments, nuclear- and electron-recoil experiments, and astrophysical
observations.

Here, we focus on collider probes for BSM-physics searches. In particular, we study high-energy
proton-proton collisions at the LHC, currently aiming to reach a center-of-mass energy of 
\SI{14}{\tera\electronvolt} in the near future. The two largest experiments at the LHC -- 
\texttt{ATLAS}~\cite{ATLAS:2008xda} and \texttt{CMS}~\cite{CMS:2008xjf} -- have hitherto mainly 
searched for events with large missing energy and/or high $p_T$ objects (jets, leptons, etc.), 
emphasizing signatures expected to stem from heavy new fields. 

Among various signatures, high-energy photons plus missing energy is one interesting 
example as it is clean with modest SM background, and is predicted in well-motivated theoretical 
models. One classic example is Gauge-mediated Supersymmetry 
Breaking (GMSB) models~\cite{Giudice:1998bp}.
Given a light and stable gravitino as the lightest supersymmetric particle (LSP), the lightest Minimal Supersymmetric Standard Model (MSSM) superpartner is actually the 
next-to-lightest supersymmetric particle (NLSP). If the NLSP is neutral, it can be either a 
neutralino, or a sneutrino. The lightest neutralino can be bino, wino, 
Higgsino, or a mixture, and can decay to a photon and a non-observable gravitino, either 
promptly or with a long lifetime; see, 
\textit{e.g.}, Refs.~\cite{Feng:2010ij,Knapen:2016exe,Kim:2017pvm,Kim:2019vcp} 
for some LHC phenomenology studies. This signature has
been searched for at the Tevatron -- at \texttt{CDF} and 
\texttt{D0}~\cite{D0:2004xvs,CDF:2013sjb} -- and at the LHC -- 
at \texttt{ATLAS}~\cite{ATLAS:2018nud} and 
\texttt{CMS}~\cite{CMS:2019zxa}.

One additional theory benchmark is a class of models with universal extra 
dimensions~\cite{Appelquist:2000nn}.
If the new dimensions are only accessible to gravity,
the lightest Kaluza-Klein particle (LKP) can decay to a photon and a gravity excitation. Both 
the lightest neutralino (assuming R-parity conservation) and the LKP should be pair-produced, 
and thus lead to the signature of two highly energetic photons plus missing energy at the LHC.

Here, we consider R-parity-violating (RPV) supersymmetry in its minimal form -- the
RPV Minimal Supersymmetric Standard Model (RPV-MSSM)~\cite{Allanach:2003eb} -- with a bino-like 
lightest neutralino as the LSP (see 
Refs.~\cite{Dreiner:1997uz,Barbier:2004ez,Mohapatra:2015fua} for reviews). The RPV-MSSM is 
as well-motivated as the  R-parity-conserving (RPC) MSSM. It not only solves the 
hierarchy problem, but also provides a natural solution to the neutrino masses 
\cite{Hall:1983id,Grossman:1998py,Hirsch:2000ef,Dreiner:2006xw,Dreiner:2011ft}, as well as a 
much richer collider phenomenology than the RPC-MSSM. In addition, it can explain various 
experimental anomalies observed in recent years, such as the 
$B$-anomalies~\cite{Trifinopoulos:2019lyo,Hu:2020yvs,Hu:2019ahp,Zheng:2021wnu,Altmannshofer:2020axr,BhupalDev:2021ipu}, 
muon $g-2$~\cite{Zheng:2021wnu,Altmannshofer:2020axr,BhupalDev:2021ipu}, and the ANITA 
anomaly~\cite{Altmannshofer:2020axr,Collins:2018jpg}.

As we discuss in more detail in \cref{sec:model} below, in the RPV-MSSM, it is possible to 
have a light neutralino of mass below \SI{10}{\giga\electronvolt}, or even massless. 
Once produced, the neutralino decays via non-vanishing RPV couplings into SM particles. Since these 
couplings are required by various (low-energy) experiments to be 
small~\cite{Allanach:1999ic,Barbier:2004ez,Bolton:2021hje}, light LSP neutralinos with mass below the $\SI{}{\giga\electronvolt}$
scale are expected to be long-lived; after production at a collider, they travel a macroscopic 
distance before decaying to SM particles.

Long-lived particles (LLPs) have in recent years received increased 
attention~\cite{Curtin:2018mvb,Lee:2018pag,Filimonova:2019tuy,Alimena:2019zri,Feng:2022inv,Schafer:2022shi}. LLPs are predicted in a wide range of BSM models such as split SUSY, RPV-SUSY, a class of portal-physics
models [axion-like particles (ALPs), heavy neutral leptons, a dark Higgs scalar, dark photons],
and models of neutral naturalness -- which are often related to 
the non-vanishing neutrino masses or dark matter. In particular, a series of dedicated 
far-detector programs have been proposed to be operated in the vicinity of LHC interaction
points (IPs), mainly aiming to look for LLPs with a proper decay length $c\tau\sim\left(1-100\right)\SI{} 
{\metre}$, or even larger. Some examples currently under discussion include \texttt{FASER}~\cite{Feng:2017uoz,FASER:2018eoc}, \texttt{FACET}~\cite{Cerci:2021nlb}, \texttt{MATHUSLA}~\cite{Chou:2016lxi,Curtin:2018mvb,MATHUSLA:2020uve}, \texttt{CODEX-b}~\cite{Gligorov:2017nwh}, \texttt{ANUBIS}~\cite{Bauer:2019vqk}, and \texttt{MoEDAL-MAPP}~\cite{Mitsou:2022vka}.

\texttt{FASER} has been approved and installed at the LHC TI12 tunnel. It consists of a small 
cylindrical decay volume of $\sim \SI{0.05}{\metre}^3$. It is expected to achieve excellent 
constraining power for a number of theoretical benchmark models such as 
ALPs~\cite{Feng:2018pew}, dark photons~\cite{Feng:2017uoz}, and inelastic dark 
matter~\cite{Berlin:2018jbm}. It is now under operation with the ongoing LHC Run 3. For the 
high-luminosity LHC (HL-LHC) period, a larger version of \texttt{FASER}, known as 
\texttt{FASER2}~\cite{FASER:2018eoc}, is also planned to be installed and running,
potentially at the same location or at a collective facility -- the Forward Physics Facility 
(FPF)~\cite{Anchordoqui:2021ghd} -- hosting various experiments, all in the very forward region 
of the LHC, including \texttt{FORMOSA}~\cite{Foroughi-Abari:2020qar} and 
\texttt{FLArE}~\cite{Batell:2021blf}. These potential future experiments are all intended to 
look for various BSM signatures.

Here, we focus on long-lived light neutralinos. They have been studied extensively for 
various present and future experiments including 
\texttt{SHiP}~\cite{deVries:2015mfw,Gorbunov:2015mba}, \texttt{ATLAS}~\cite{deVries:2015mfw}, 
far detectors at the LHC~\cite{Helo:2018qej,Dercks:2018eua,Dercks:2018wum,Dreiner:2020qbi},
\texttt{Belle~II}~\cite{Dey:2020juy}, 
\texttt{Super-Kamiokande}~\cite{Candia:2021bsl}, and future lepton 
colliders~\cite{Wang:2019orr,Wang:2019xvx}. These works mostly consider the signature of 
a neutralino decay into a charged lepton plus a meson, induced by $LQ\bar{D}$ operators~\cite{Domingo:2022emr}, while 
the production can result from decays of either mesons, $\tau$ leptons, or $Z$-boson. 

In this work, we propose a novel signature associated with very light lightest-neutralino $\left(\tilde{\chi}^0_1\right)$ decays: A single photon plus missing energy. Such a signature can appear as a result of the radiative decay associated with neutrinos, 
\begin{equation}
	\tilde{\chi}^0_1\to \nu_i+\gamma \text{ or } \bar{\nu}_i + \gamma\,,
\end{equation}
arising at the loop level via the RPV couplings $\lam'_{ijj}$ of the
$LQ\bar{D}$ operators or $\lam_{ijj}$ of the $LL\bar{E}$ operators. This 
decay can dominate in certain mass ranges and for certain choices of 
RPV couplings.\footnote{We note that light long-lived particles (LLPs)
	decaying to a light neutrino and a photon may explain the \texttt{MiniBooNE}
	anomaly~\cite{Fischer:2019fbw,Brdar:2020tle}, but given the recent 
	negative results by \texttt{MicroBooNE}~\cite{MicroBooNE:2021rmx}, and 
	possible SM explanations for the 
	anomaly~\cite{Giunti:2019sag,Ioannisian:2019kse}, we do not 
	consider it any further here.} We consider the lightest neutralino to be produced from rare decays of mesons such as pions and $B$-mesons
copiously created at the LHC, and study the probing potential of \texttt{FASER} and \texttt{FASER2} to these scenarios, for the 
signature of a single, displaced photon. As discussed in \cref{sec:simulation}, the background is expected to be negligible.

The paper is organized as follows. We briefly introduce the RPV-MSSM, as well as the light neutralino scenario in the next section. In \cref{sec:scenarios} we present a list of representative benchmark scenarios, which we investigate in this paper. In \cref{sec:experiment} we discuss the experimental setup at \texttt{FASER} and \texttt{FASER2}, and in \cref{sec:simulation} we detail our simulation procedure for estimating the sensitivity reach. The results are then presented with a discussion in \cref{sec:results}. We conclude the paper with a summary and an outlook in \cref{sec:conclusions}.

\section{Theoretical Framework}\label{sec:model}
Here, we discuss the underlying supersymmetric model, as well as details of the
light neutralino scenario.

\subsection{The R-parity-violating MSSM}
Given the $\left(N=1\right)$ supersymmetry algebra, and the MSSM particle content, 
the most general $\mathrm{SU}(3)_{C}\times\mathrm{SU}(2)_{L}\times\mathrm{U}(1)_Y$-invariant, renormalizable superpotential can be written as,
\begin{equation}
	W = W_{\mathrm{MSSM}} + W_{\mathrm{LNV}} + W_{\mathrm{BNV}}\,,
	\label{eq:TFeq1}
\end{equation}
where $W_{\mathrm{MSSM}}$ is the usual MSSM superpotential -- see, for instance,
Ref.~\cite{Allanach:2003eb} -- while the terms, 
\begin{eqnarray}
	W_{\mathrm{LNV}} = \frac{1}{2}\lam^{ijk}L_iL_j\Bar{E}_k + 
	\lam'^{ijk}L_iQ_j\Bar{D}_k + \kappa^{i}H_uL_i\,, \;\qquad 
	W_{\mathrm{BNV}} = 
	\frac{1}{2}\lam''^{ijk}\Bar{U}_i\Bar{D}_j\Bar{D}_k\,,
	\label{eq:TFeq1a}
\end{eqnarray}
violate lepton- and baryon-number, respectively. In the above, $L$ $(Q)$, and 
$\Bar{E}$ $(\Bar{U},\Bar{D})$ are the MSSM lepton (quark) $\mathrm{SU}(2)_{L}$-doublet and $\mathrm{SU}(2)_{L}$-singlet chiral superfields, respectively, while $H_u, H_d$ label the 
$\mathrm{SU}(2)_{L}$-doublet Higgs chiral superfields. We do not show gauge 
indices explicitly but write the generational ones: $i,j,k=1,2,3$ with a summation 
implied over repeated indices. The $\lam$'s are dimensionless coupling parameters, 
the $\kap$'s are dimension-one mass parameters.

The combined lepton- and baryon-violation contained in the above terms
may cause the proton to decay too 
quickly~\cite{Chamoun:2020aft,ParticleDataGroup:2020ssz}. Thus, in the
MSSM, all operators in $W_{\mathrm{LNV}} + W_{\mathrm{BNV}}$ are set to
zero by invoking a $\mathbb{Z}_2$ symmetry called 
R-parity~\cite{Farrar:1978xj}. This allows $W_{\mathrm{MSSM}}$ while
disallowing $W_{\mathrm{RPV}}\equiv W_{\mathrm{LNV}}+ W_{\mathrm 
	{BNV}}$. However, the proton can be protected without completely 
forbidding $W_{\mathrm{RPV}}$. For instance, forbidding $W_{\mathrm 
	{BNV}}$, while keeping $W_{\mathrm{LNV}}$, results in a stable proton. 
Baryon triality -- $B_3$ -- is a $\mathbb{Z}_3$-symmetry that achieves 
exactly 
this~\cite{Dreiner:2006xw,
	Ibanez:1991pr,Dreiner:2005rd,Dreiner:2012ae}.

Importantly, RPV phenomenology can be starkly different compared to the
RPC case~\cite{Dreiner:1991pe,Dreiner:1997uz,Barbier:2004ez,deCampos:2007bn,Dercks:2017lfq}. 
The LSP is no longer guaranteed to be stable leading to vastly different final state signatures. 
The collider phenomenology of RPV models is rich and 
complex~\cite{Dreiner:1991pe,Dercks:2017lfq}, and it is crucial that our SUSY 
search strategies cover all possibilities. We now discuss in some detail one 
interesting realization of RPV-SUSY: A very light neutralino.

\subsection{A Very light Lightest-Neutralino}

In principle, any supersymmetric particle can be the LSP in RPV 
models~\cite{Dercks:2017lfq,Dreiner:2008ca,Desch:2010gi}. Here, we
restrict ourselves to the case of a neutralino. Potentially important mass
bounds come from colliders, dark matter (cosmology), and astrophysics.
For collider searches of a stable neutralino, the strongest bound comes
from LEP, $m_{\tilde\chi^0_1}\gtrsim\SI{46}{\giga 
	\electronvolt}$~\cite{ParticleDataGroup:2020ssz}. This is based on
chargino searches, and assumes the grand-unified mass relation is 
satisfied between the electroweak supersymmetry breaking gaugino masses, 
$M_1 = \frac{5}{3}\tan^2\theta_W M_2 \approx 0.5 M_2$, with $\theta_W$ the electroweak mixing angle. However, once the
relation is dropped, the mass of the lightest neutralino is experimentally unconstrained~\cite{Dreiner:2009ic}.
Such a scenario typically requires the lightest neutralino to be dominantly 
bino-like~\cite{Choudhury:1999tn,Dreiner:2009ic}.

A stable lightest neutralino is further constrained by dark matter 
limits. The Lee-Weinberg bound gives $m_{\tilde\chi^0_1}\gtrsim 
\mathcal{O}\left(10\right)\SI{}{\giga\electronvolt}$
\cite{Lee:1977ua,Belanger:2001am,Hooper:2002nq,Bottino:2002ry,Dreiner:2003wh,Dreiner:2009ic,AlbornozVasquez:2010nkq,Calibbi:2013poa,KumarBarman:2020ylm}. 
However, in RPV models where the LSP is unstable, this bound does not
apply~\cite{Dreiner:2009ic}.

Then, from our discussion above, if the RPV couplings are small -- which is what one expects -- 
the neutralino can be stable on collider scales while unstable on 
cosmological scales, thus evading all existing constraints. 
Such a neutralino is allowed to be very light and, in principle, even
massless~\cite{Dreiner:2009ic,ParticleDataGroup:2020ssz}. It is
also consistent with astrophysical constraints, such as the cooling of supernoavae 
and white dwarfs, if the sfermions are heavy enough~\cite{Dreiner:2003wh,Kachelriess:2000dz,Dreiner:2013tja}.

We next consider the phenomenology of RPV-SUSY scenarios with such light neutralinos as the LSP. If the neutralino is massive enough, and/or the RPV couplings are
sizeable, such that the proper decay length of the neutralino is $c\tau \lsim\mathcal 
{O}\left(1\right)\SI{}{\metre}$, various RPV searches performed at \texttt{ATLAS} and 
\texttt{CMS} -- including those for displaced vertices -- apply; see, \textit{e.g.}, Refs.~\cite{ATLAS:2020uer,CMS:2021tkn}. These searches rely on detecting the decay 
products of the neutralino, which can contain jets and leptons, depending on 
the dominant RPV couplings. On the other hand, for very light neutralinos,
and/or if the RPV couplings are very small, the neutralino LSP is stable on 
macroscopic scales. Then, the signature is invisible to colliders, just as in the RPC 
case. Thus, as long as heavier SUSY particles are produced at the LHC, that then
cascade-decay down to the neutralino LSP, the RPC searches for large missing 
transverse momentum apply even to the RPV case.

However, in light of to-date unsuccessful supersymmetry searches, one
possibility is that the heavier SUSY spectrum may be inaccessible at 
the LHC. Very light neutralinos, $m_{\tilde\chi^0_1} \lsim\mathcal{O} 
\left(4.5\right)\SI{}{\giga\electronvolt}$, can still be produced in 
abundance in such a scenario in RPV models through the rare decays of
mesons via an $LQ\bar D$ operator~\cite{Choudhury:1999tn,deVries:2015mfw,Dedes:2001zia}. These 
neutralinos would be highly boosted in the forward direction of the momentum of the decaying meson. None of
the above search strategies applies in such a case, and the scenario represents a 
realistic possibility of low-scale SUSY manifesting in a way that would have escaped 
our searches so far. With the long-lived particle programs at the LHC picking up pace, 
there is the possibility of filling this gap. If the highly boosted, light neutralino 
decays with a proper decay length, $c\tau\sim\mathcal{O}\left(1-100\right)\SI{} 
{\metre}$, it may be visible in dedicated far-detector experiments such as \texttt{FASER}.
Before we discuss the decay modes of such light neutralinos, we provide, for completeness, the unpolarized decay width of pseudoscalar mesons into a light neutralino and a lepton via an $LQ\bar{D}$ operator, reproduced from Ref.~\cite{deVries:2015mfw},

\begin{eqnarray}\label{eqn:production_rate_neutralino}
\Gamma\left(M_{a b} \rightarrow \widetilde{\chi}_1^0+l_i\right)=\frac{\lambda^{\frac{1}{2}}\left(m_{M_{a b}}^2, m_{\widetilde{\chi}_1^0}^2, m_{l_i}^2\right)}{64 \pi m_{M_{a b}}^3}\left|G_{i a b}^{S, f}\right|^2\left(f_{M_{a b}}^S\right)^2\left(m_{M_{a b}}^2-m_{\widetilde{\chi}_1^0}^2-m_{l_i}^2\right)\,,
\end{eqnarray}
where $l_i$ denotes a charged lepton $\ell_i^{\pm}$ or a neutrino $\nu_i$, depending on whether $M_{a b}$ is charged or neutral, and $\lambda^{\frac{1}{2}}$ is the K\"all\'en function $\lambda^{\frac{1}{2}}(x, y, z) \equiv \sqrt{x^2+y^2+z^2-2 x y-2 x z-2 y z}$. The coupling constants $G_{i a b}^{S, f}$ and the meson decay constant $f_{M_{a b}}^S$ are defined as in Ref.~\cite{deVries:2015mfw}.
In particular, the $LQ\bar{D}$ coupling $\lambda'$ is proportional to $G_{i a b}^{S, f}$.
In the above, the charge-conjugated mode is implied.

\subsection{Neutralino Decay}

\begin{figure}[h]
	\centering
	\includegraphics[width=0.49\textwidth]{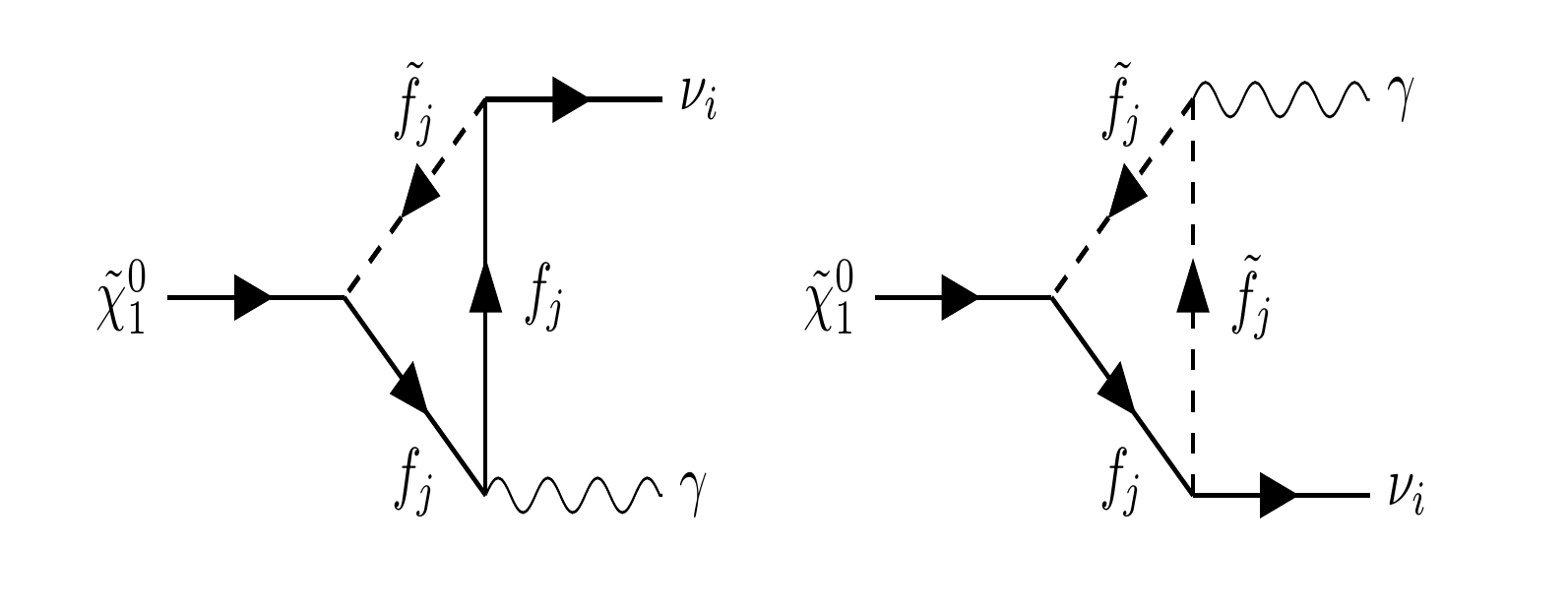}
	\caption{Feynman diagrams for the radiative neutralino decay.} 
	\label{fig:rad-neutralino-decay-feynman}
\end{figure}

The dominant decay mode of the neutralino is dictated by the relative sizes of the RPV
couplings, as well as the neutralino mass~\cite{Domingo:2022emr}. For $m_{\tilde\chi^ 
	0_1} \lsim\mathcal{O}\left(4.5\right)\SI{}{\giga\electronvolt}$, the neutralino can
decay into a meson and a lepton via an $LQ\bar D$ operator, if kinematically allowed. 
Similarly, it can decay as $\tilde\chi^0_1\to\ell^+\ell'^-\nu + \text{c.c.}$ via
the $LL\bar E$ operators. For operators $L_iQ_j\bar D_j$ or $L_iL_j\bar E_j$, there is
also the possibility for the loop-induced decays, 
\begin{equation}
	\tilde\chi^0_1\to \left(\gamma +\nu_i\,,\gamma +\bar\nu_i\right)\,,
	\label{eq:rad-neutralino-decay-reaction}
\end{equation}
which has essentially no kinematic threshold.
We show example Feynman diagrams in~\cref{fig:rad-neutralino-decay-feynman}. The 
fermions/sfermions in the loop have generation index $j$. The decay rate is given 
by~\cite{Hall:1983id,Dawson:1985vr,Haber:1988px,Domingo:2022emr}:  
\begin{align}
	\Gamma(\tilde\chi^0_1\to \gamma +\nu_i)&=
	\frac{\lambda^2\alpha^2m^3_{\tilde{\chi}^0_1}} {512\pi^3\cos^2\theta_W}\left[\sum_f 
	\frac{e_f N_c m_f\left(4e_f+1\right)}{m^2_{\tilde{f}}} \left(1+\log{\frac{m^2_f} 
		{m^2_{\tilde{f}}}}\right)\right]^2\label{eq:radiative-neutralino-decay}\\
	&=\Gamma\left(\tilde\chi^0_1\to \gamma +\bar\nu_i\right)\,.\nonumber
\end{align}
In the above expression, $\lambda$ is the relevant $L_iQ_j\bar D_j$ 
or $L_iL_j\bar E_j$ coupling, $\alpha$ is the (QED) fine-structure
constant, while $\theta_W$ is the electroweak mixing angle. $e_f, N_c$
and $m_f(m_{\tilde{f}})$ are the electric charge in units of $e$,
color factor ($3$ for $LQ\bar D$, $1$ for $LL\bar E$), and the mass,
respectively, of the fermion (sfermion) inside the loop.
We note that the above simple formula for the width neglects any mixings in the scalar sector. While this effect -- depending on the supersymmetric parameters -- may become significant, it introduces several undetermined SUSY-parameters in the expression. At the level of precision of our study, we find it convenient to work with this simplified approximation.
The two decay widths in~\cref{eq:radiative-neutralino-decay} are equal as a result of the Majorana nature of the neutralino. The logarithmic function in Eq.~\eqref{eq:radiative-neutralino-decay}, log$\frac{m_f^2}{m^2_{\tilde{f}}}$, changes only by about a factor of two if we vary $m_{\tilde{f}}$ between \SI{1}{\tera\electronvolt} and \SI{100}{\tera\electronvolt}.
Therefore, in our numerical simulations, we will fix $m_{\tilde{f}}$ at \SI{1}{\tera\electronvolt} for the log term, so that we can use $\lambda/m^2_{\tilde{f}}$ as a single combined parameter, without separating $\lambda$ and $m_{\tilde{f}}$.

Despite the loop-suppression, the radiative mode,~\cref{eq:rad-neutralino-decay-reaction}, can be relevant for very light 
neutralinos. The partial width is proportional to $\left(m^3_{\tilde{\chi}^0_1}m^2_f 
\right)/m^4_{\tilde f}$, compared to $m^5_{\tilde{\chi}^0_1}/m^4_{\tilde f}$ for the tree-level three-body decay into fermions 
\cite{Hall:1983id,Dawson:1985vr,Dreiner:1991pe}, and can thus be
important for small masses. Depending on the generation indices of the
dominant RPV coupling(s) and the neutralino mass, it might even be 
the only kinematically allowed mode. 
In this paper we focus on the scenario where the neutralino dominantly
decays as in~\cref{eq:rad-neutralino-decay-reaction}, as this channel has not been considered 
before in the context of long-lived light neutralino searches. We now present some benchmark scenarios for phenomenological studies.

\section{Benchmark Scenarios}\label{sec:scenarios}
In order to study the phenomenology of a very light neutralino decaying only via the radiative mode,~\cref{eq:rad-neutralino-decay-reaction}, we
present some representative benchmark scenarios which we believe
cover all relevant possibilities, and which we investigate in detail 
in the next section. We list the corresponding parameters in 
\cref{tab:benchmarks-rad-neutralino}. In each case, we assume the
listed couplings are the only non-negligible RPV couplings. The 
neutralino is produced through the rare decay of the meson $M$ via 
the coupling $\lam^{\text{P}}_{ijk}$: $M\to\tilde\chi^0_1+\ell\left( 
\nu\right)$
\cite{Choudhury:1999tn,Dedes:2001zia,Dreiner:2009er,deVries:2015mfw}, 
and then decays in one of the ways discussed in the previous section 
via the coupling $\lam^\text{D}_{ijj}$.\footnote{We note that, in this
	work, we are neglecting the effects of the suppressed three-body decay 
	that can proceed at one-loop level via an off-shell $Z$, e.g.,
	$\tilde\chi^0_1\to 3\nu\,$. We thank Florian Domingo for a 
	discussion on this topic.} In the table, we also list the current best 
bounds on the couplings $\lam^\text {P}_{ijk}$ and $\lam^\text{D}_{ijj}$.

For benchmark \textbf{B1}, the neutralino is produced via the most 
abundant mesons at the LHC -- pions -- in association with muons (neutrinos). This occurs via the coupling $\lam^\text{P}_{ijk} 
=\lam'_{211}$. The charged production mode ($\pi^\pm\to\tilde\chi^0_1+\mu^\pm$) is 
only possible if the mass of the neutralino satisfies the bound, 
\begin{equation}
	m_{\tilde\chi^0_1} < m_{\pi^{\pm}}-m_{\mu^{\pm}}\approx 
	\SI{35}{\mega\electronvolt}\,.
\end{equation}
For neutralinos heavier than the above threshold, only the neutral production mode ($\pi^0\to\tilde\chi^0_1+\nu_\mu$) contributes; however, this mode is suppressed owing to the short lifetime of the neutral pion which translates into a low decay branching fraction into neutralinos.
For the benchmark, we choose ${m_{\tilde 
		\chi^0_1}}=\SI{30}{\mega\electronvolt}$; the lightness of the neutralino means that the radiative mode is the only kinematically allowed decay. In principle, with the coupling $\lam^\text{P}_{ijk} = \lam'_{111}$ 
instead, a heavier neutralino can be produced in charged pion 
decays: $\pi^\pm\to \tilde\chi^0_1+e^\pm$, but the severe bound, \cite{Allanach:1999ic}
\begin{equation}
	\lambda'_{111} \lsim 0.001 \left(\frac{m_{
			\tilde{d}_R}}{\SI{1}{\tera\electronvolt}}\right)^2\,,
\end{equation}
implies this mode can not be probed at the experiments we consider 
here. 

For the decay coupling, we choose $\lam^\text{D}_{ijj}=\lam' 
_{333}$. The decay width,~\cref{eq:radiative-neutralino-decay}, 
is roughly proportional to $m_f^2$. Thus, the heavier the fermion in the 
loop, the shorter the lifetime of the neutralino. For the very light 
neutralino in {\bf B1}, we require a heavy fermion in the loop to get
testable scenarios at \texttt{FASER}; we expect maximum sensitivity to 
couplings $\lam'_{i33}$ or $\lam_{i33}$.

\begin{table}[t]
	\scalebox{0.85}{\begin{tabular}{ccccc}
			\toprule \toprule
			{\bf Scenario} & $\mathbf{m_{\tilde\chi^0_1}}$ & {\bf Production} $\left(\lambda^\text{P}_{ijk}\right)$ 
			& {\bf Decay} $\left(\lambda^\text{D}_{ijj}\right)$ 
			& {\bf Current Constraints}\\
			{\bf B1} & $\SI{30}{\mega\electronvolt}$ & $\lambda'_{211} \left(M=\pi^\pm, \pi^0\right)$ & $\lam'_{333}$ & $\lam'_{211} < 0.59 \left(\frac{m_{
					\tilde{d}_R}}{\SI{1}{\tera\electronvolt}}\right),\; \lambda'_{333} < 1.04$\\
			{\bf B2} & $\SI{75}{\mega\electronvolt}$ & $\lambda'_{212}\left(M=K^\pm, K^0_{L/S}\right)$ & $\lambda'_{333}$ & $\lambda'_{212} < 0.59 \left(\frac{m_{
					\tilde{s}_R}}{\SI{1}{\tera\electronvolt}}\right),\;  \lambda'_{333} < 1.04$\\
			{\bf B3} & $\SI{200}{\mega\electronvolt}$ & $\lambda'_{112}\left(M=K^\pm, K^0_{L/S}\right)$ & $\lambda_{322}$ & $\lambda'_{112} < 0.21 \left(\frac{m_{
					\tilde{s}_R}}{\SI{1}{\tera\electronvolt}}\right),\; \lambda_{322} < 0.7 \left(\frac{m_{
					\tilde{\mu}_R}}{\SI{1}{\tera\electronvolt}}\right)$\\
			{\bf B4} & $\SI{300}{\mega\electronvolt}$ & $\lambda'_{221}\left(M=D^\pm,K^0_{L/S}\right)$ & $\lambda_{233}$ & $\lambda'_{221} < 1.12\,,\; \lambda_{233} < 0.7 \left(\frac{m_{
					\tilde{\tau}_R}}{\SI{1}{\tera\electronvolt}}\right)$\\
			{\bf B5} & $\SI{500}{\mega\electronvolt}$ & $\lambda'_{222}\left(M=D_S^\pm\right)$ & $\lambda'_{222}$ & $ \lambda'_{222} < 1.12$\\[1.5mm]
			{\bf B6} & $\SI{1}{\giga\electronvolt}$ & $\lambda'_{313}\left(M=B^\pm, B^0\right)$ & $\lambda'_{333}$ & $\lambda'_{313} < 1.12\,,\; \lambda'_{333} < 1.04$\\
			\bottomrule \bottomrule
	\end{tabular}}
	\caption{Benchmark scenarios considered in this paper. The neutralino is produced through
		the rare decay of the meson $M$ via the coupling $\lam^\text {P}_{ijk}$: $M\to\tilde 
		\chi^0_1+\ell\left(\nu\right)$. The neutralino decay is as 
		in~\cref{eq:rad-neutralino-decay-reaction} via the coupling $\lam^\text{D}_{ijj}$. The 
		photon energy in the neutralino rest frame is $E_\gamma=m_{\tilde\chi^0_1}/2$, but can 
		range from $\mathcal{O}\left(0.1\right)$ to $\mathcal{O}\left(1\right)\SI{}{\tera\electronvolt}$ at \texttt{FASER}. In the furthest-to-the-right column, we list the current best bounds on the couplings, see for example, Ref.~\cite{Allanach:1999ic}.} 
	\label{tab:benchmarks-rad-neutralino}
\end{table}

For the benchmarks \textbf{B2} and \textbf{B3}, we choose the parameters such that the neutralinos are produced in kaon decays. This time, unlike the pion case, both the charged and neutral modes have comparable contributions. For \textbf{B2}, 
the neutralino decays only radiatively, as 
in~\cref{eq:rad-neutralino-decay-reaction}. For \textbf{B3}, the decay 
coupling $\lam^{\mathrm{D}}=\lam_{322}$ also allows for tree-level 
leptonic decays:
\begin{align}
	\tilde\chi^0_1\to \left(\nu_\tau\mu^\pm\mu^\mp,\,\tau^\pm\mu^\mp\nu_\mu
	\right) + \text{c.c.}\,.
\end{align} 
However, these are kinematically blocked for $m_{\tilde\chi^0_1}\lsim2
m_\mu$. Thus, we have chosen $m_{\tilde\chi^0_1}=\SI{200}{\mega\electronvolt}$. Later, when 
we present numerical results, we go beyond the strict parameters in the benchmark scenarios and consider plots in the RPV coupling vs.~neutralino mass plane. One then has to account for the fact that additional decay modes can open. Note that in \textbf{B3}, we now select a decay 
coupling that is not third generation in the last two indices, since 
the neutralino is now heavy enough to avoid a too-small decay width, even for lighter fermions in the loop.

We have chosen benchmark \textbf{B4} such that a single coupling leads to production of the neutralinos from 
both kaons and $D^{\pm}$. Since kaons are more abundant at the LHC than $D$-mesons, the former production mode contributes more to the neutralino flux. For the selected mass of $\SI{300}{\mega\electronvolt}$, there are no other relevant decay modes of the neutralino than the radiative one.
But for the coupling vs.~mass plot, the neutralino can decay into kaons 
above the relevant thresholds. The neutralino \textit{production} 
through kaons is, of course, blocked for these heavier masses.
In addition, for this scenario and the ones below, there can also be three-body decays into two mesons and a lepton, mediated via the $LQ\bar D$ operators; these can become relevant in the very high mass regime, $m_{\tilde\chi^0_1}\gtrsim 
\mathcal{O}\left(1.5\right)\SI{}{\giga\electronvolt}$.
These are neglected for simplicity in the present work as their impact for sub-GeV neutralinos -- which are the focus of our study -- is minor.
We will also neglect any Cabibbo-Kobayashi-Maskawa-mixing effects for similar reasons.
More details on these effects can be found in Ref.~\cite{Domingo:2022emr}.

Benchmark \textbf{B5} is a special case: It 
requires only a single non-zero RPV coupling ($\lam'_{222}$) for both production (via $D_S^\pm$ mesons) and decay. This is absent in the tree-level neutralino decay case~\cite{deVries:2015mfw}, except for an extremely small mass-window of around \SI{4}{\mega\electronvolt}. For the given mass,
$m_{\tilde\chi^0_1}=\SI{500}{\mega\electronvolt}$, the neutralino decays only radiatively. 
But at higher masses, it may decay into $\eta, \eta',$ or $\phi$.

Finally, we have chosen benchmark \textbf{B6} such that the neutralinos are produced via $B$-meson decays, thereby allowing the neutralino to be relatively heavy, leading to more energetic photons. The neutralino is produced in association with
a $\tau^{\pm}\left(\nu_\tau \right)$ via the charged (neutral) mode; the two modes have comparable contributions. For
$m_{\tilde\chi^0_1}>m_{B^\pm}-m_{\tau^{\pm}}$, only the neutral mode is
kinematically allowed. The radiative mode is the only relevant decay channel.

We note in passing the interesting observation that the radiative decay of a neutralino gives us a method of producing significant $\nu_
{\tau}$ fluxes. These are suppressed in the SM. With \texttt{FASER$\nu$}~\cite{FASER:2019dxq,FASER:2020gpr}
under operation, this may give us an interesting opportunity to detect the
neutralino by looking for $\nu_\tau$ events. However, we leave an 
investigation in this direction for the future.

\begin{figure}[h!]
	\centering
	\includegraphics[width=0.495\textwidth]{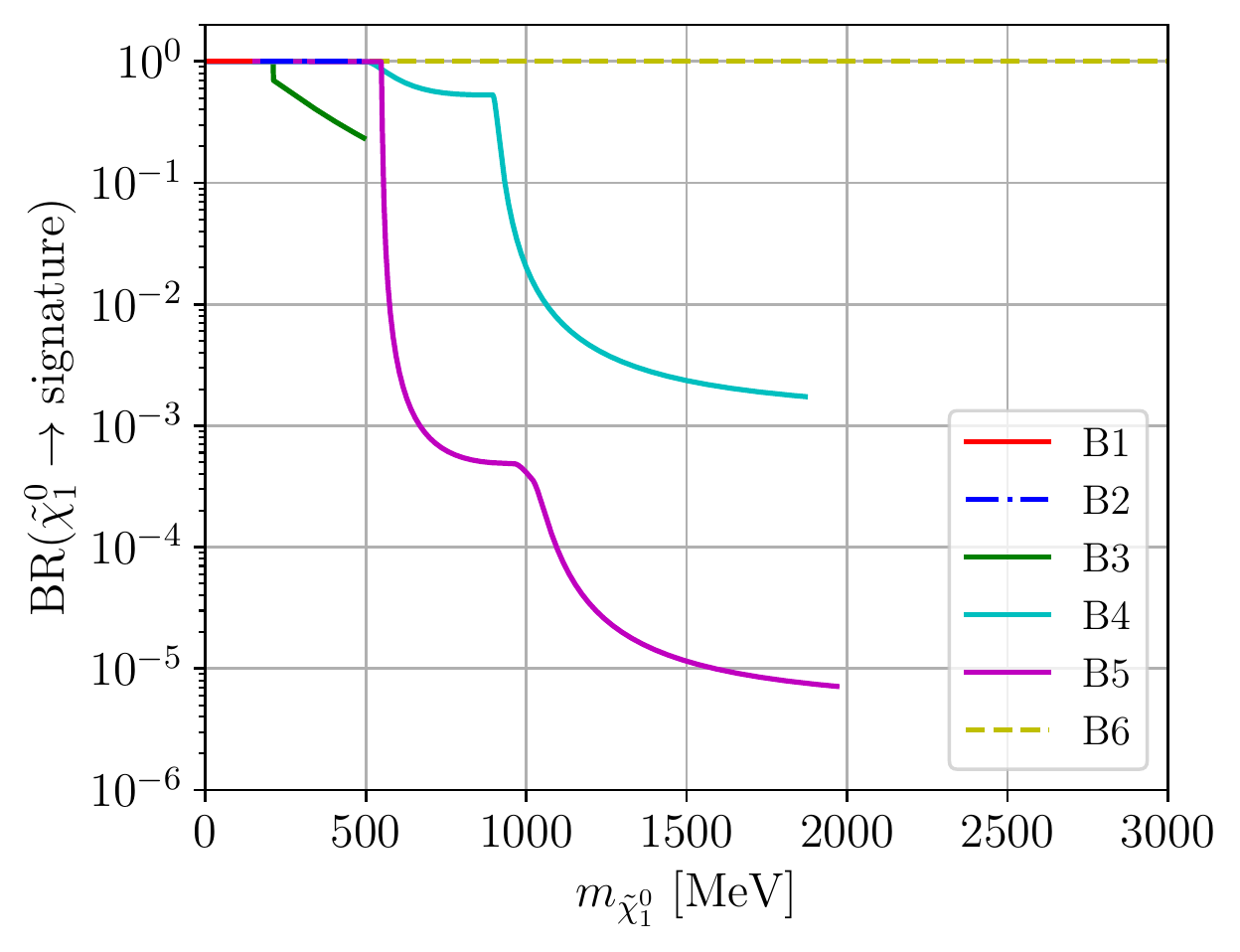}
	\caption{Branching ratios of the lightest neutralino into the single-photon signature, with varying neutralino mass.}
	\label{fig::BRsig_vs_mass}
\end{figure}

Before closing the section, we provide a plot in~\cref{fig::BRsig_vs_mass}, showing the decay branching ratios of the lightest neutralino into our signature, $\gamma+\overset{\scriptscriptstyle(-)}{\nu}$, as a function of the neutralino mass, for all the considered benchmark scenarios.

\section{\texttt{FASER} Experiment}\label{sec:experiment}

The \texttt{FASER} experiment~\cite{Feng:2017uoz,FASER:2018eoc} is 
a cylindrical detector that has recently been installed inside the TI12 
tunnel, $\SI{480}{\metre}$ from the \texttt{ATLAS} IP along the beam collision axis line of
sight. The detector is composed of tracking stations, scintillators, and
a calorimeter. The  cylinder axis is along the extended beam collision 
axis. Its decay volume has a radius of $\SI{10}{\centi\metre}$ and a length of $\SI{1.5}{\metre}$. It
is currently running during Run 3 of the LHC and is expected to collect data from proton-proton collisions of around $\SI{150}{\femto\barn^{-1}}$
integrated luminosity. At the front end of \texttt{FASER}, an additional
emulsion detector known as
\texttt{FASER$\nu$}~\cite{FASER:2019dxq,FASER:2020gpr} has been installed, which is aimed at detecting high-energy neutrinos produced at the \texttt{ATLAS} IP. In this work, we do not study the potential of \texttt{FASER$\nu$}.

A follow-up experiment -- \texttt{FASER2}~\cite{FASER:2018eoc} -- is 
currently slated to be operated during the HL-LHC period. If it is to be installed at the same position as \texttt{FASER}, it will be at a distance of $\SI{480}{\metre}$ from the \texttt{ATLAS} IP. Otherwise, it could be one of 
the experiments to be hosted at the FPF~\cite{Anchordoqui:2021ghd}, 
$\SI{620}{\metre}$ from the \texttt{ATLAS} IP, also along the beam axis. We expect the difference
between $\SI{480}{\metre}$ and $\SI{620}{\metre}$ distance to the IP to lead to only relatively minor changes in the sensitivity reach, as discussed in 
Ref.~\cite{Anchordoqui:2021ghd}. For this study, we work with the 
geometrical setup of a radius of $\SI{1}{\metre}$ and a length of $\SI{5}{\metre}$ for the 
\texttt{FASER2} decay fiducial volume, and consider it $\SI{480}{\metre}$ away from 
the IP. By the end of Run 5 at the LHC, \texttt{FASER2} should have
collected about $\SI{3}{\atto\barn^{-1}}$ integrated luminosity of collision data.
Similarly, an emulsion detector has been proposed to be installed 
at the front face of \texttt{FASER2}, known as \texttt{FASER$\nu$2}.
We will assume the detector components of \texttt{FASER2} and hence 
detection principles and efficiencies are similar to those of 
\texttt{FASER}, except for the different geometrical acceptances.

In Ref.~\cite{Feng:2018pew}, the authors studied 
axion-like particles (ALPs) at \texttt{FASER}, where the ALPs decay to 
\textit{a pair} of photons. They estimated that the calorimeter
spatial resolution should be sufficiently good for resolving the two photons with an efficiency of about 50\%, and the background should be 
negligible for diphoton events. Here, our signature includes only
\textit{one} photon. To provide a discussion on the expected background 
level, we follow the arguments given in Ref.~\cite{Jodlowski:2020vhr}. At \texttt{FASER}, the single photons are detected as high-energy deposits 
in the electromagnetic calorimeter. Other objects may also cause such deposits, \textit{e.g.}, neutrinos 
interacting deep inside the calorimeter via charged-current interactions. In order to differentiate the photon signal, a pre-shower station has been installed right before the calorimeter~\cite{FASER:2021cpr} which first converts the photon, thereby identifying it.
Moreover, during Run 3 of the LHC program, the \texttt{FASER} detector is planned to be upgraded with a high-precision preshower detector. This would allow to distinguish two very closely spaced highly energetic photons~\cite{Boyd:2803084}.
Furthermore, neutrinos and muons coming from the IP can penetrate the $\SI{100}{\metre}$ of rock in
front of \texttt{FASER} and reach the detector with energies in the $\SI{}{\tera\electronvolt}$
scale. These neutrinos could interact with the detector resulting in
energetic particles including individual photons. However, these
energetic photons are accompanied by tens of charged particles, allowing
to veto such events easily with the tracker stations. The muons could
radiate high-energy photons as well, mainly via Bremsstrahlung, but the 
veto stations positioned right in front of the \texttt{FASER} decay 
volume~\cite{FASER:2021cpr} should enable the rejection of muon-associated
events.\footnote{One possible background that we neglect here could come from off-axis muons that can penetrate \texttt{FASER} without passing through the veto stations; estimating such a background would require a detailed simulation. We thank Max Fieg for bringing up this point.}
Finally, neutral pions produced in hadronic showers initiated by muons in the absorber-rock material could also constitute a background for our signal if the two photons produced in their decays cannot be spatially resolved or only one of them is observed.\footnote{We thank Michael Albrow for bringing this to our attention.} In such a case, requiring an energy threshold for the signal may help since the photons from our signal are expected to be more energetic; see Ref.~\cite{Jodlowski:2020vhr} for more details on the point of using energy thresholds.
A detailed estimate, however, requires a full simulation of hadronic interactions inside the rock.
In this work, we will assume zero background for our signal.

Since the search proposed here with the single-photon signature does not
require the usage of the tracker, in principle the tracker volume could 
be considered as effectively part of the fiducial volume. Taking this
into account would allow to enlarge the length of the fiducial volume of
\texttt{FASER} and \texttt{FASER2} by roughly $\SI{1}{\metre}$~\cite{FASER:2021cpr}
and $\SI{5}{\metre}$~\cite{Anchordoqui:2021ghd}, respectively, enhancing the 
sensitivity reach to some extent. In this work, we only comment on this
possibility and choose to stay with the standard benchmark geometries,
as given explicitly above.

There are several other past and ongoing experiments that should have sensitivity to a
radiatively decaying light neutralino. These include beam-dump experiments such as 
\texttt{LSND}~\cite{LSND:1996jxj}, \texttt{E613}~\cite{Ball:1980ojt}, 
\texttt{MiniBooNE}~\cite{MiniBooNE:2008hfu}, \texttt{E137}~\cite{Bjorken:1988as}, and 
\texttt{NA64}~\cite{Gninenko:2016kpg,Gninenko:2013rka,NA64:2017vtt}, as well as
$B$-factory experiments such as \texttt{BaBar}~\cite{BaBar:2001yhh} and 
\texttt{Belle~II}~\cite{Belle-II:2010dht,Belle-II:2018jsg}. Typically, each of these 
experiments is optimized to primarily produce only a certain type of meson, at rates which
could be higher than the LHC. Correspondingly, they can probe a subset of the RPV models 
we have presented here in a somewhat cleaner environment. The LHC has the advantage of 
producing all types of mesons at significant rates, thus providing a
scenario-independent probe. However, given a signal, it could be difficult to disentangle
the underlying model(s). Further, given the different center-of-mass energies (and hence 
the spectra of the produced neutralinos), and the detector layouts, the phase-space 
region probed by these other experiments may complement that probed by \texttt{FASER}. However, detailed 
simulations are required to make more precise statements; this is beyond the scope of the present work. 

Further, limits coming from searches for heavy neutral leptons can also be relevant for us. We will include these in our plots in~\cref{sec:results}.

Finally, we note that our signature could also be probed by \texttt{FACET} -- a proposed new subsystem of the \texttt{CMS} experiment.
In this study, however, we only focus on \texttt{FASER} and \texttt{FASER2}.

\section{Simulation}\label{sec:simulation}

We now proceed to describe the simulation procedure for estimating the number
of signal events in the two experiments. We use the package 
\texttt{FORESEE}~\cite{Kling:2021fwx} to obtain the neutralino spectrum 
in the far-forward region, relevant for \texttt{FASER} and 
\texttt{FASER2}. As mentioned, the dominant sources of the neutralinos
are the rare decays of mesons produced at the \texttt{ATLAS} IP:
\begin{equation}
	M\to\tilde\chi^0_1+\ell\left(\nu\right)\,.
	\label{eq:meson-decay-reaction}
\end{equation}
Direct pair-production of neutralinos, in comparison, is expected to be several
orders of magnitude lower~\cite{Helo:2018qej,Dedes:2001zia,deVries:2015mfw}, and is hence neglected here. We
include all possible production modes for the different benchmark scenarios, summing over all meson contributions, to estimate the total number of produced neutralinos over the runtime of the experiment. However, it is
necessary but not sufficient for the mother meson to decay into a neutralino: 
The meson itself may be long-lived, \textit{e.g.}, charged pions and kaons. 
Thus, we require the meson to decay before hitting any absorber material or 
leaving the beam pipe; otherwise, the meson could be stopped and the neutralino 
is no longer boosted in the direction of \texttt{FASER}. Keeping this in mind, we use \texttt{FORESEE} to determine the neutralino production rate and spectrum from the meson spectrum by specifying the decay branching ratios corresponding to~\cref{eq:meson-decay-reaction}. The generated spectrum is two-dimensional, in terms of angle and momentum.

\begin{table}[t!]
	\begin{center} 
		\renewcommand{\arraystretch}{1.2}
		\begin{tabular}{cccccc} 
			\hline \hline
			{\bf Detector} &  $\mathbf{\mathcal{L}}$ & $\mathbf{\sqrt{s}}$  &  $\mathbf{L}$ & 
			$\mathbf{\Delta}$ &   $\mathbf{R}$\\
			{\bf \texttt{FASER}} & $\SI{150}{\femto\barn^{-1}}$ & $\SI{14}{\TeV}$ & $\SI{480}{\meter}$ & $\SI{1.5}{\meter}$ & 
			$\SI{10}{\centi\meter}$ \\
			{\bf \texttt{FASER2}} & $\SI{3000}{\femto\barn^{-1}}$ & $\SI{14}{\TeV}$ & $\SI{480}{\meter}$ & $\SI{5}{\meter}$ & 
			$\SI{1}{\meter}$ \\
			\hline
			\hline
		\end{tabular}
	\end{center}
	\caption{Integrated luminosities and geometries of the detectors used in
		the simulations. Here, $\mathbf{\mathcal{L}}$, $\mathbf{\sqrt{s}}$, $\mathbf{L}$, 
		$\mathbf{\Delta}$, and $\mathbf{R}$ label, respectively, the integrated 
		luminosity, the collider center-of-mass energy, the distance from the IP,
		the detector length, and the detector radius.}
	\label{tab:detector_simulation}
\end{table}

We also use \texttt{FORESEE} to compute the probability for the 
neutralino to decay inside the detector volume. See~\cref{tab:detector_simulation} for the values
corresponding to the detector position and geometry we employ in our
simulation for \texttt{FASER} and \texttt{FASER2}. We take into account the 
full neutralino lifetime, $\tau_{\tilde \chi^0_1}$, as well as its kinematics. The former is computed using all possible decay channels of the neutralino (including the decay into pseudoscalar and vector mesons) as a function of its mass and the non-vanishing RPV couplings, \textit{cf.} the discussion in~\cref{sec:scenarios}. However, in the numerical results presented in the next section, we have chosen an explicit signal for 
detecting the neutralino decay. Although all neutralino decays inside the
detector are technically visible, we estimate the signal strength based on the specific radiative mode alone. This is done
to avoid the consideration of background events; the decay into a neutrino and a photon gives a clean and unique 
signature.

Given the neutralino spectra, we
estimate the number of decays that occur inside the detector defined by 
its position and geometry.  For the analysis, the simulation 
takes into account the distance L between the \texttt{ATLAS} IP and the \texttt{FASER} detector, and the acceptance rate $P[\tilde{\chi}_1^0]$ in terms of the neutralino's three-momentum, its position of production (accounting for the mesons' lifetimes), as well as the lifetime of the neutralino itself. In our simulation, we do not make any momentum cuts. By further specifying the branching ratio into the radiative mode, the simulation counts the number of signal events passing the selection criteria. Thus, we can finally 
estimate the number of single-photon neutralino decay observations,
\begin{align}
	N^\text{obs}_{\tilde{\chi}_1^0} = P[\tilde{\chi}_1^0] \cdot  
	\text{BR}\left[\tilde{\chi}_1^0 \to \left(\gamma +\nu_i\,,\gamma +\bar\nu_i\right) \right]\cdot 
	\sum_{\text{mesons}} N^{\text{prod}}_{\tilde{\chi}_1^0}\,.
\end{align}
We stress again that we assume zero background, \textit{cf.} the discussion in~\cref{sec:experiment}. Further, we assume a detector efficiency of 100\%.

\begin{figure}[h!]
	\centering
	\includegraphics[width=0.495\textwidth]{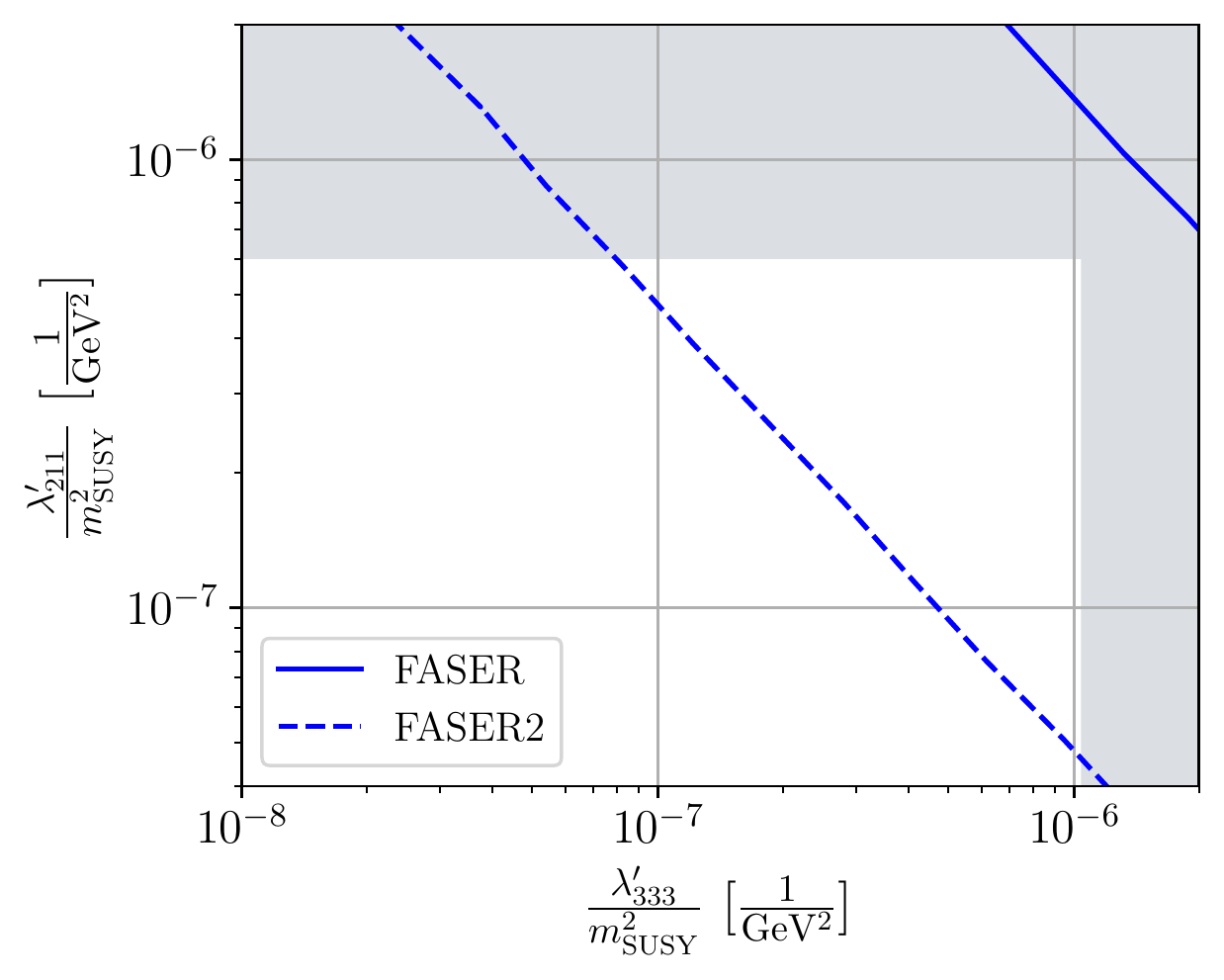}
	\includegraphics[width=0.495\textwidth]{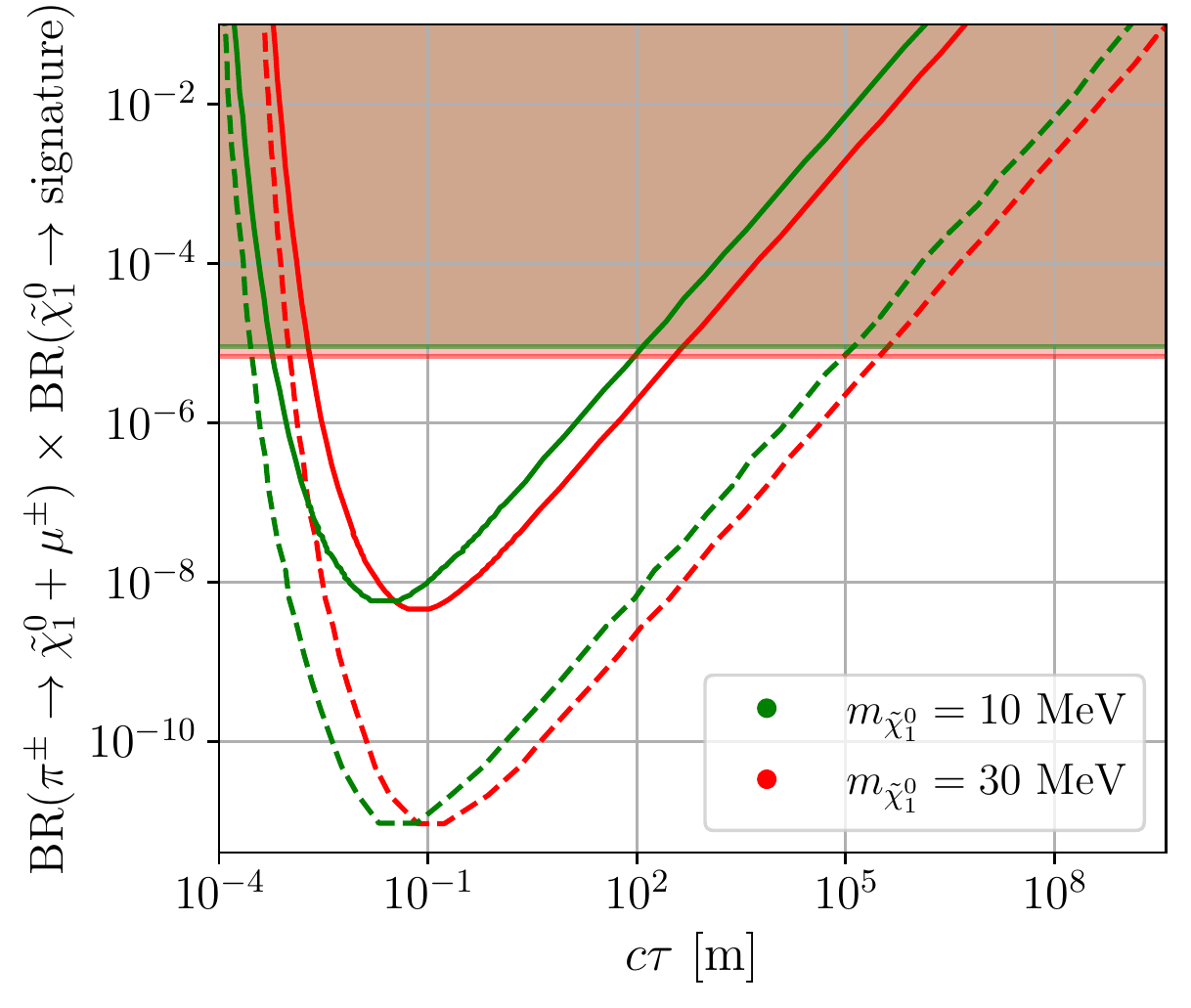}
	\caption{Sensitivity reach for \texttt{FASER} (solid lines) and \texttt{FASER2} (dashed 
		lines) for the benchmark scenario \textbf{B1}, \textit{cf.}~\cref{tab:benchmarks-rad-neutralino}.
		The left plot shows the sensitivity reach in the production coupling ($\frac{\lam^{\prime}_{211}}{m_\text{SUSY} 
			^{2}}$) vs.~decay coupling ($\frac{\lam^ 
			{\prime}_{333}}{m_\text{SUSY}^{2}}$) plane, for a neutralino mass of \SI{30}{\mega\electronvolt}.
		The gray areas are excluded by the low-energy 
		bounds, also given in~\cref{tab:benchmarks-rad-neutralino}. The right plot shows
		the sensitivity reach in BR$(\pi^\pm\to\tilde{\chi}^0_1+\mu^\pm)\times$BR$(\tilde{\chi}^0_1\to\text{signature})$ as a function of the 
		neutralino decay length, $c\tau$, for $m_{\tilde\chi^0_1}=10$ and $\SI{30}{\mega 
			\electronvolt}$. The shaded regions correspond to existing constraints from HNL searches.}
	\label{fig::llp_neutralino::couplingvcoupling1}
\end{figure}

\begin{figure}[h!]
	\centering
	\includegraphics[width=0.495\textwidth]{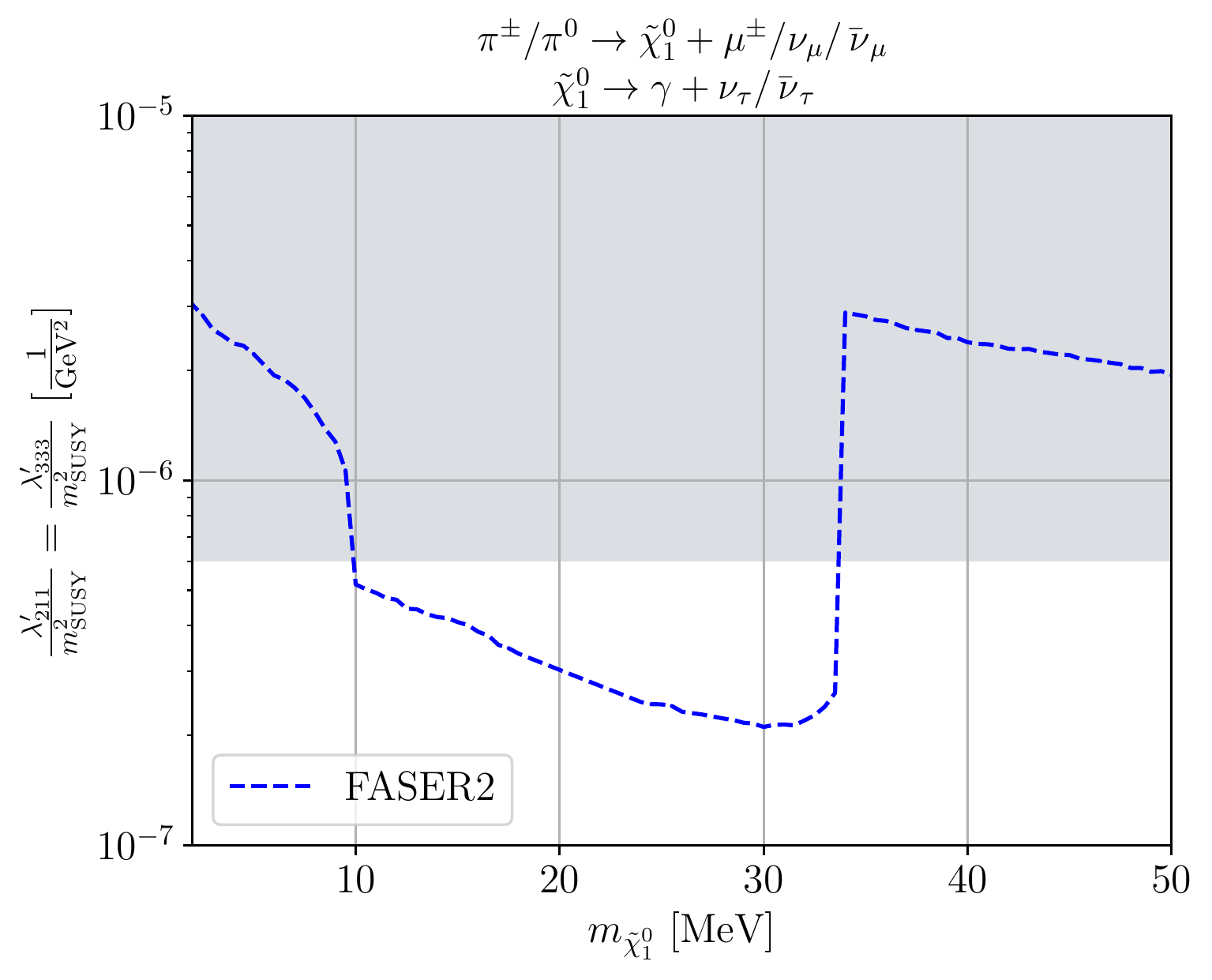}
	\caption{Sensitivity reach in the neutralino mass-coupling plane for \texttt{FASER2} for the same physics scenario as in
		\textbf{B1} but with variable  neutralino mass. The production ($\frac{\lambda^{\prime}_{211}}{m^2_\text{SUSY}}$) 
		and decay ($\frac{\lambda^{\prime}_{333}}{m^2_\text{SUSY}}$) couplings have been set equal. The gray areas are 
		excluded by the low-energy bounds.}
	\label{fig::llp_neutralino::couplingvmass1}
\end{figure}

\section{Numerical Results}\label{sec:results}

We now present our numerical results. For the sensitivity limits, we 
require the observation of 3 radiative decays of the lightest neutralino
in the detector for an integrated luminosity at the LHC of $\SI{150} 
{\femto \barn^{-1}}$ for \texttt{FASER}, and $\SI{3000}{\femto\barn 
	^{-1}}$ for \texttt{FASER2}. This corresponds to a potential 95\,\% 
confidence-level exclusion limit under the assumption of vanishing 
background. 

We first show, in~\cref{fig::llp_neutralino::couplingvcoupling1}, 
results for the benchmark scenario \textbf{B1} of 
\cref{tab:benchmarks-rad-neutralino}. On the left, we plot the 
sensitivity in the $\lam^{\text{P}}/m^2_{\text{SUSY}}=\lam^\prime_{211}
/m^2_{\text{SUSY}}$ versus $\lam^{\text{D}}/m^2_{\text{SUSY}}=\lam^ 
\prime_{333}/m^2_{\text{SUSY}}$ plane for a fixed neutralino mass of \SI{30}{\mega\electronvolt}.
In gray we include the low-energy bounds given in~\cref{tab:benchmarks-rad-neutralino}, for fixed sfermion masses of \SI{1}{\tera\electronvolt} (the same choice is taken for the other model-dependent plots in this section).
We see that \texttt{FASER} has no new sensitivity for this scenario beyond the 
low-energy bounds, whereas \texttt{FASER2} can extend the reach by more
than an order of magnitude in $\lam^{\mathrm{P}}/m^2_{\text{SUSY}}$ 
or $\lam^{\mathrm{D}}/m^2_{\text{SUSY}}$ in units of $\SI{}{\giga\electronvolt}^{-2}$. The
right plot in~\cref{fig::llp_neutralino::couplingvcoupling1} is model-independent, in that it is valid for any new, neutral long-lived particle 
(LLP) produced in charged pion decays, which decays with a signature at \texttt{FASER} or 
\texttt{FASER2}, here specifically with a mass of $10$ or $\SI{30}{\mega 
	\electronvolt}$. The maximum sensitivity (the minima of the 
curves) depends on the location of the detector, and also 
on the momentum distribution of the produced pions and, correspondingly,
of the pions' decay product neutralinos~\cite{Dercks:2018eua}. That 
is why the minimum of the curve shifts to slightly
smaller LLP lifetimes for lighter LLP masses, which are more boosted. We
see that \texttt{FASER} (\texttt{FASER2}) can probe the product of the decay branching fractions of the charged pion into an LLP and a muon and the LLP into the signature, down to a few times $10^{-9}$ 
$\left(10^{-12}\right)$.
We note that existing searches for heavy neutral leptons (HNLs), $N$, which mix with active neutrinos and are produced from pion decays, may be recast into bounds on the right plot.
The leading bounds for HNLs of mass \SI{10}{\mega\electronvolt} and \SI{30}{\mega\electronvolt} in $\pi^\pm \to \mu^\pm + N$ decays stem from two peak searches: Ref.~\cite{Daum:1987bg}, and Ref.~\cite{PIENU:2019usb}, respectively.
The former shows a bound of $10^{-5}$ on BR$(\pi^\pm \to \mu^\pm + N)$ for mass \SI{10}{\mega\electronvolt}. Ref.~\cite{PIENU:2019usb} presents $90\%$ confidence-level exclusion limits in the mixing-squared vs.~mass plane; we convert these into limits on BR$(\pi^\pm \to \mu^\pm+N)$~\cite{Gorbunov:2020rjx}, obtaining a bound of $6.9\times 10^{-6}$ for mass \SI{30}{\mega\electronvolt}. These two bounds are model-independent and are plotted as shaded areas in the right plot of~\cref{fig::llp_neutralino::couplingvcoupling1}, using BR$(\tilde{\chi}^0_1 \to \text{signature})=1$.
One easily observes that \texttt{FASER} and \texttt{FASER2} are sensitive to large parts of the parameter space beyond these existing bounds.

\begin{figure}[htb]
	\centering
	\includegraphics[width=0.495\textwidth]{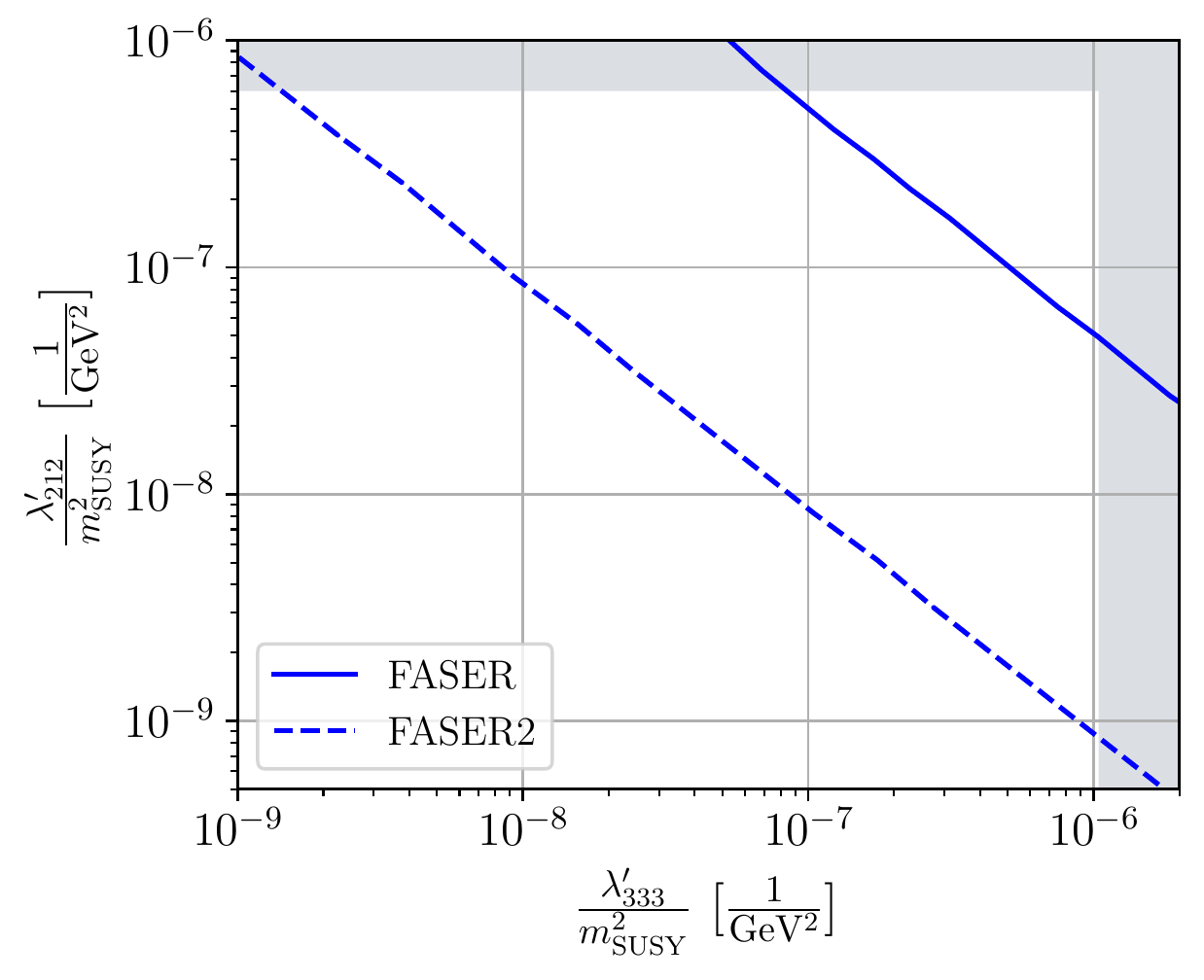}
	\includegraphics[width=0.495\textwidth]{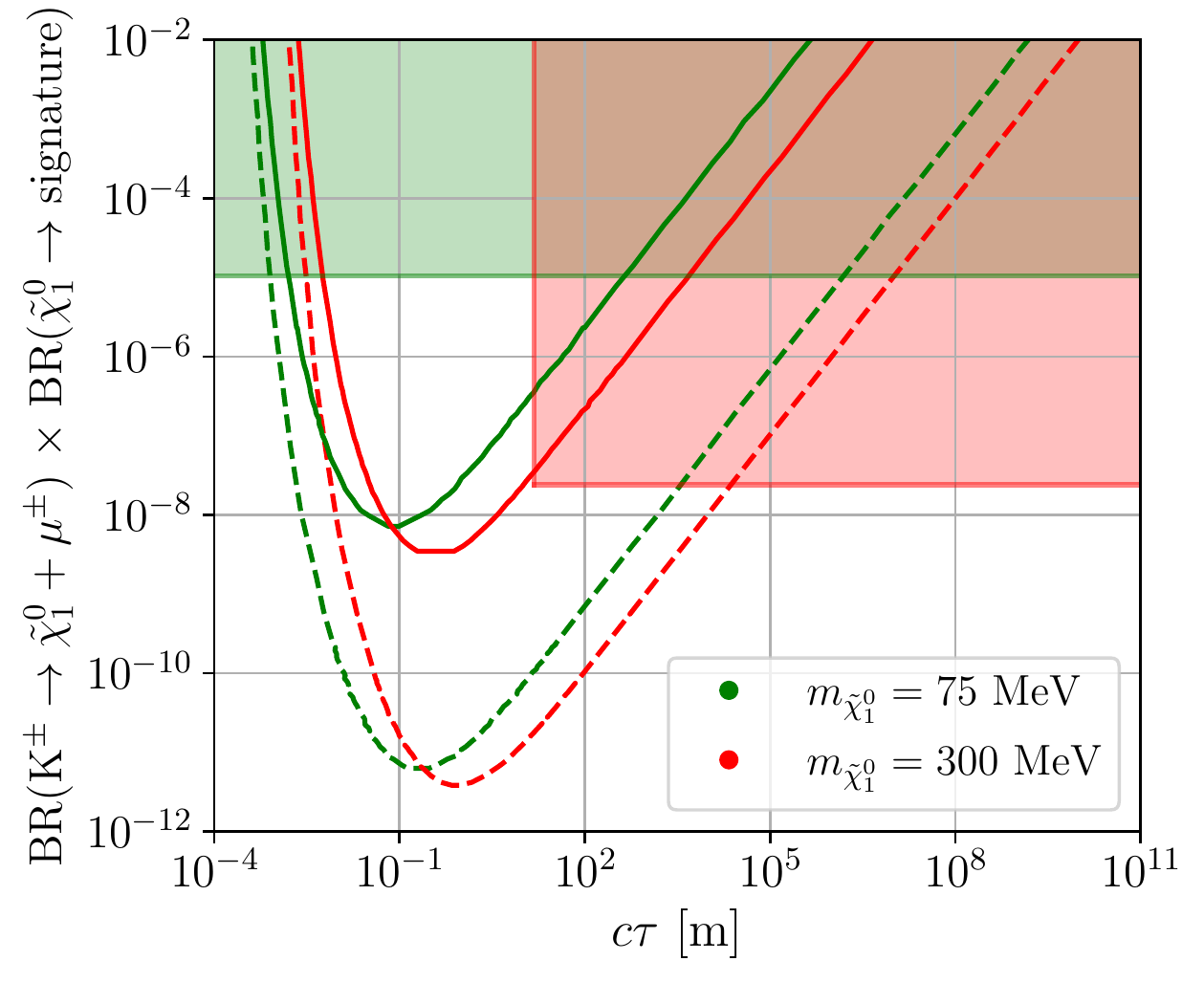}
	\caption{As in~\cref{fig::llp_neutralino::couplingvcoupling1} but for the benchmark 
		scenario \textbf{B2} with $m_{\tilde{\chi}_1^0}=\SI{75}{\mega\electronvolt}$, \textit{cf}.~\cref{tab:benchmarks-rad-neutralino}. The right plot shows the
		sensitivity reach in BR$(K^\pm\to\tilde {\chi}^0_1+\mu^\pm)\times$BR$(\tilde{\chi}^0_1\to\text{signature})$ as a function of the neutralino decay length, $c\tau$, for $m_{\tilde\chi^0_1}=75$ and $\SI{300}{\mega\electronvolt}$.}
	\label{fig::llp_neutralino::couplingvcoupling2}
\end{figure}

\cref{fig::llp_neutralino::couplingvmass1} shows the sensitivity reach
of \texttt{FASER2} for the benchmark scenario \textbf{B1} 
of~\cref{tab:benchmarks-rad-neutralino}, but allowing the neutralino 
mass to vary and fixing $\lam^P=\lam^D$. The gray band, as before, 
indicates the low-energy constraints on the couplings. Since 
\texttt{FASER} does not provide any new sensitivity reach beyond these 
low-energy bounds, we do not depict it in the plot. We see a maximum
sensitivity is reached for neutralino masses between $10$ and 
\SI{35}{\mega\electronvolt}.  In general, the sensitivity reach in the couplings improves; for instance, as we increase the neutralino mass up to \SI{30}{\mega\electronvolt}, and again in the region beyond 
\SI{35}{\mega\electronvolt}. Heavier neutralinos translate into shorter 
lifetimes, and hence more decays of the neutralino within the volume of 
\texttt{FASER2}, \textit{cf.}~\cref{eq:radiative-neutralino-decay}. 
There is, however, a sharp drop in sensitivity near the neutralino mass, 
$m_{\tilde\chi^0_1}\sim \SI{34}{\mega\electronvolt}$. This is the
threshold for the decay of charged pions to neutralinos accompanied by a muon. The branching 
fraction of the neutral pion mode, $\pi^0 \to\tilde \chi^0_1+\nu_\mu$, is suppressed by 
the short lifetime.

\begin{figure}[htb]
	\centering
	\includegraphics[width=0.495\textwidth]{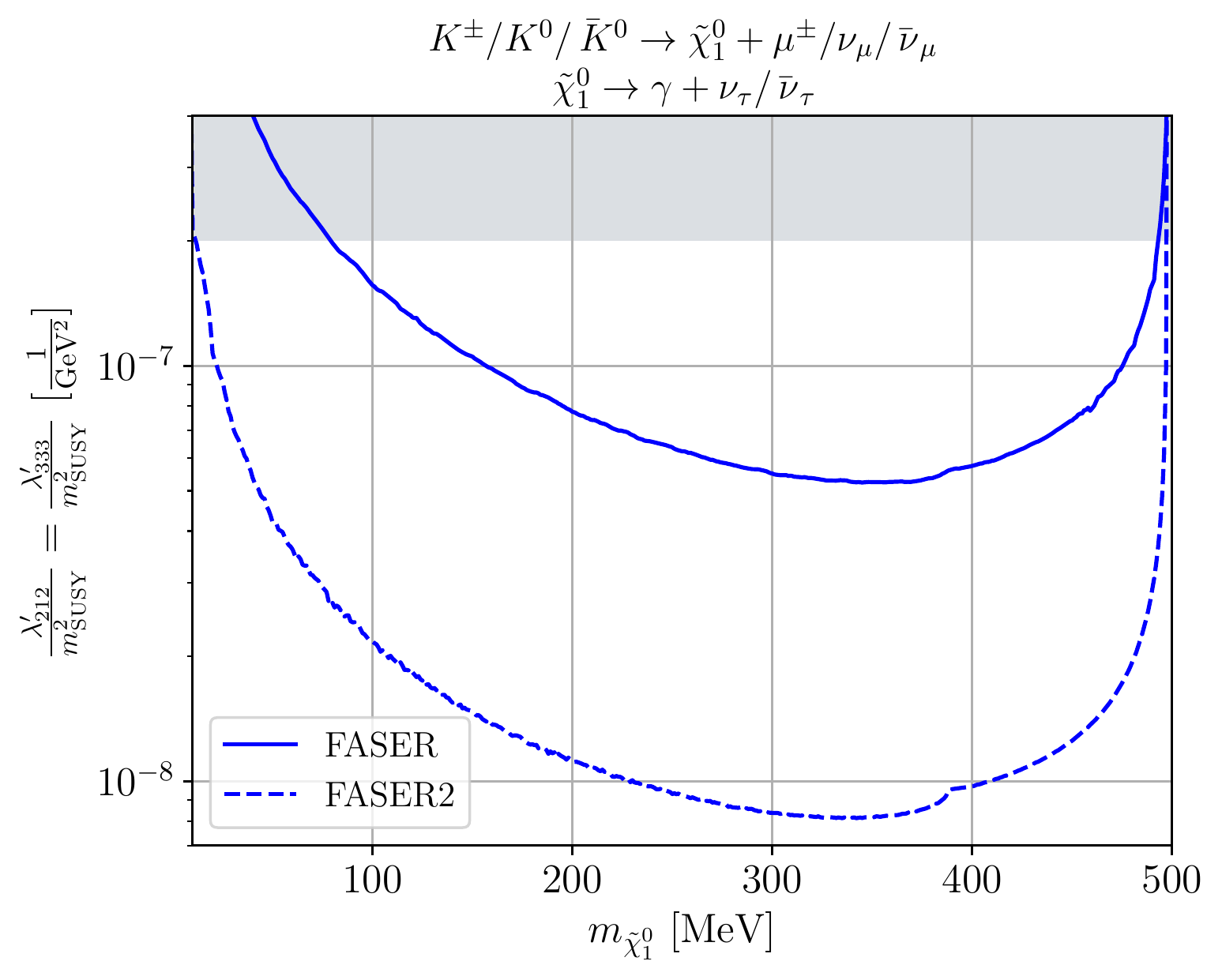}
	\includegraphics[width=0.495\textwidth]{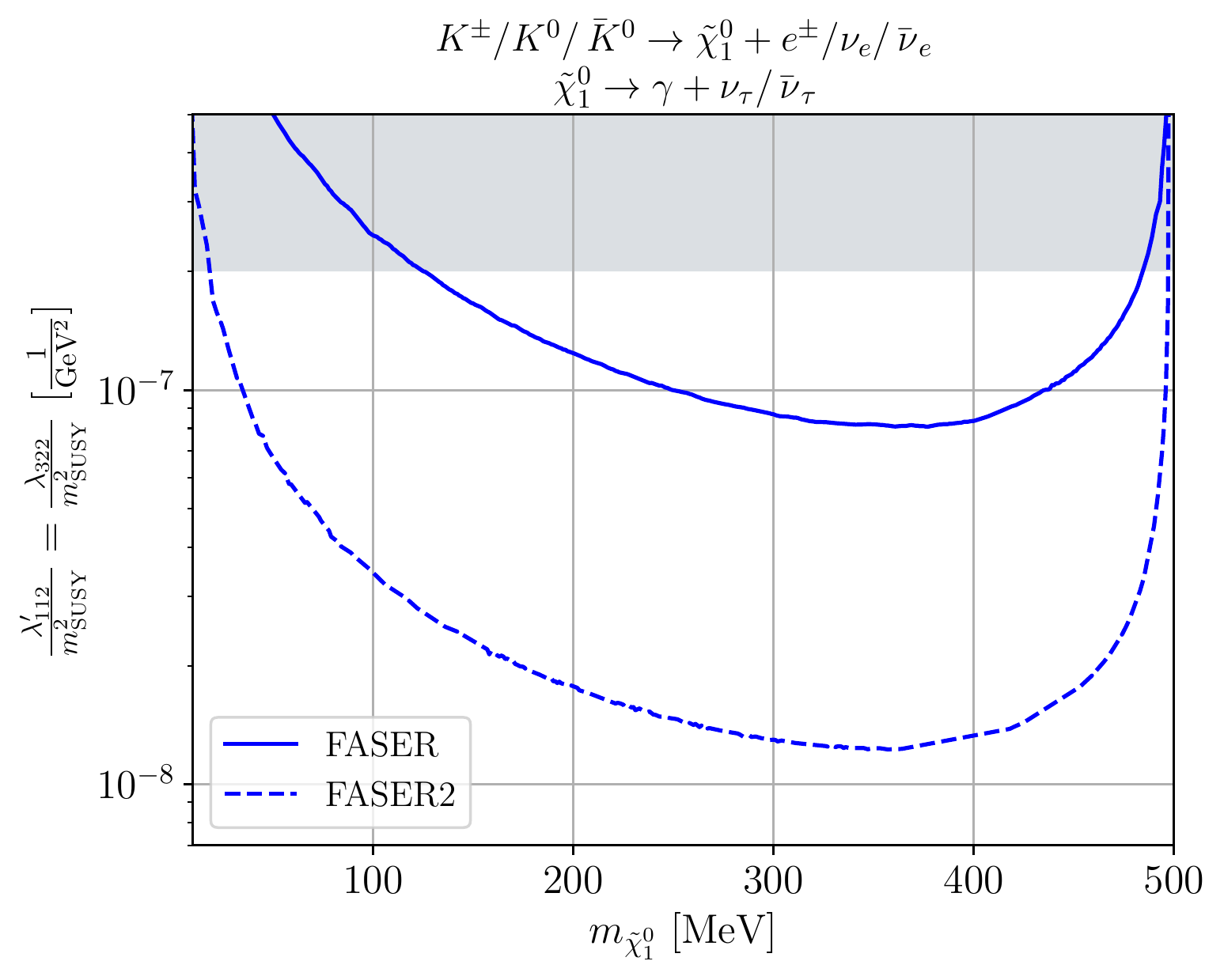}
	\caption{As in~\cref{fig::llp_neutralino::couplingvmass1} but for the
		benchmark scenarios \textbf{B2} (left) and \textbf{B3} (right). The sensitivity reach corresponds to \texttt{FASER} 
		(solid line) and \texttt{FASER2} (dashed line).}
	\label{fig::llp_neutralino::couplingvmass2}
\end{figure}

\begin{figure}[htb]
	\centering
	\includegraphics[width=0.495\textwidth]{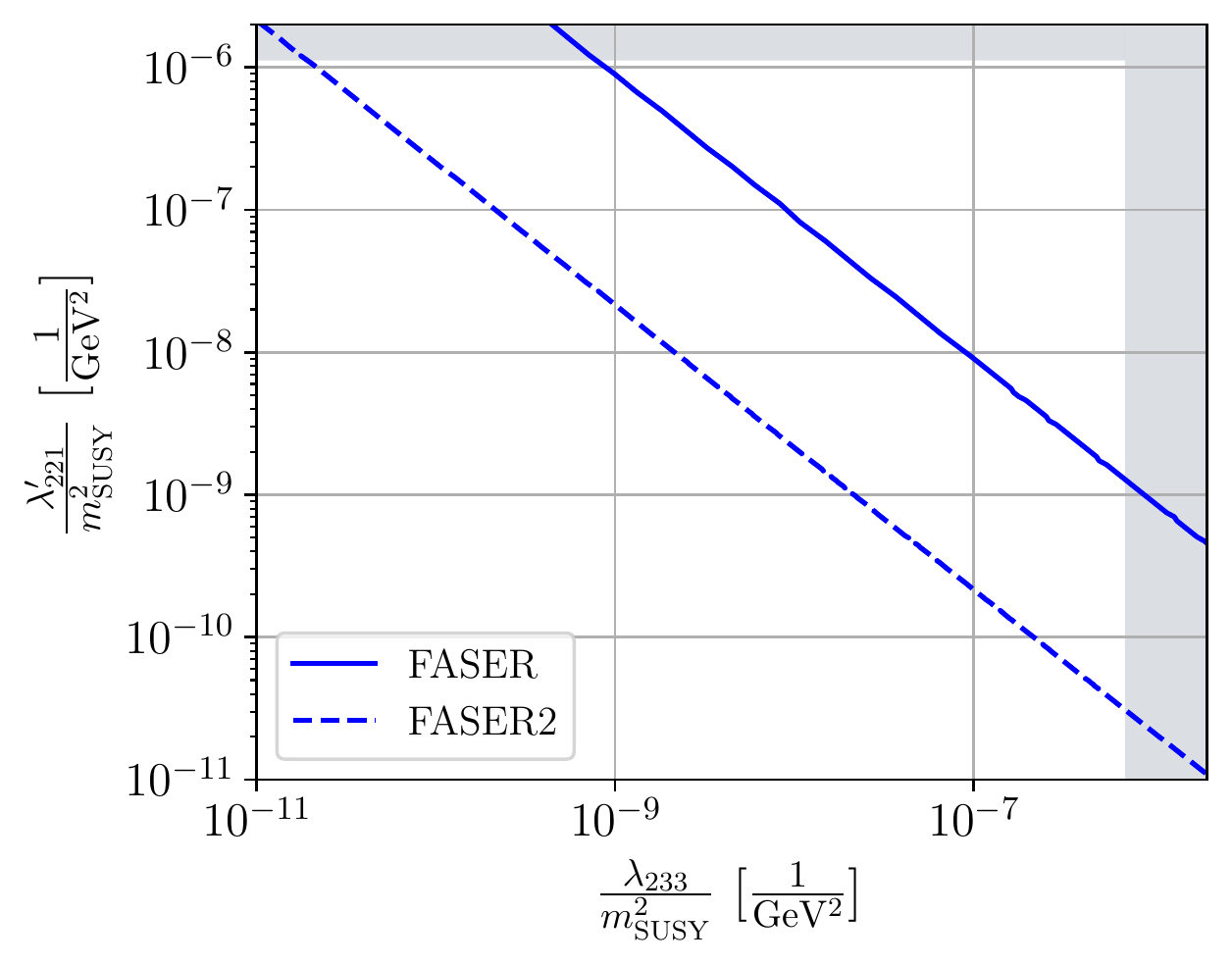}
	\includegraphics[width=0.495\textwidth]{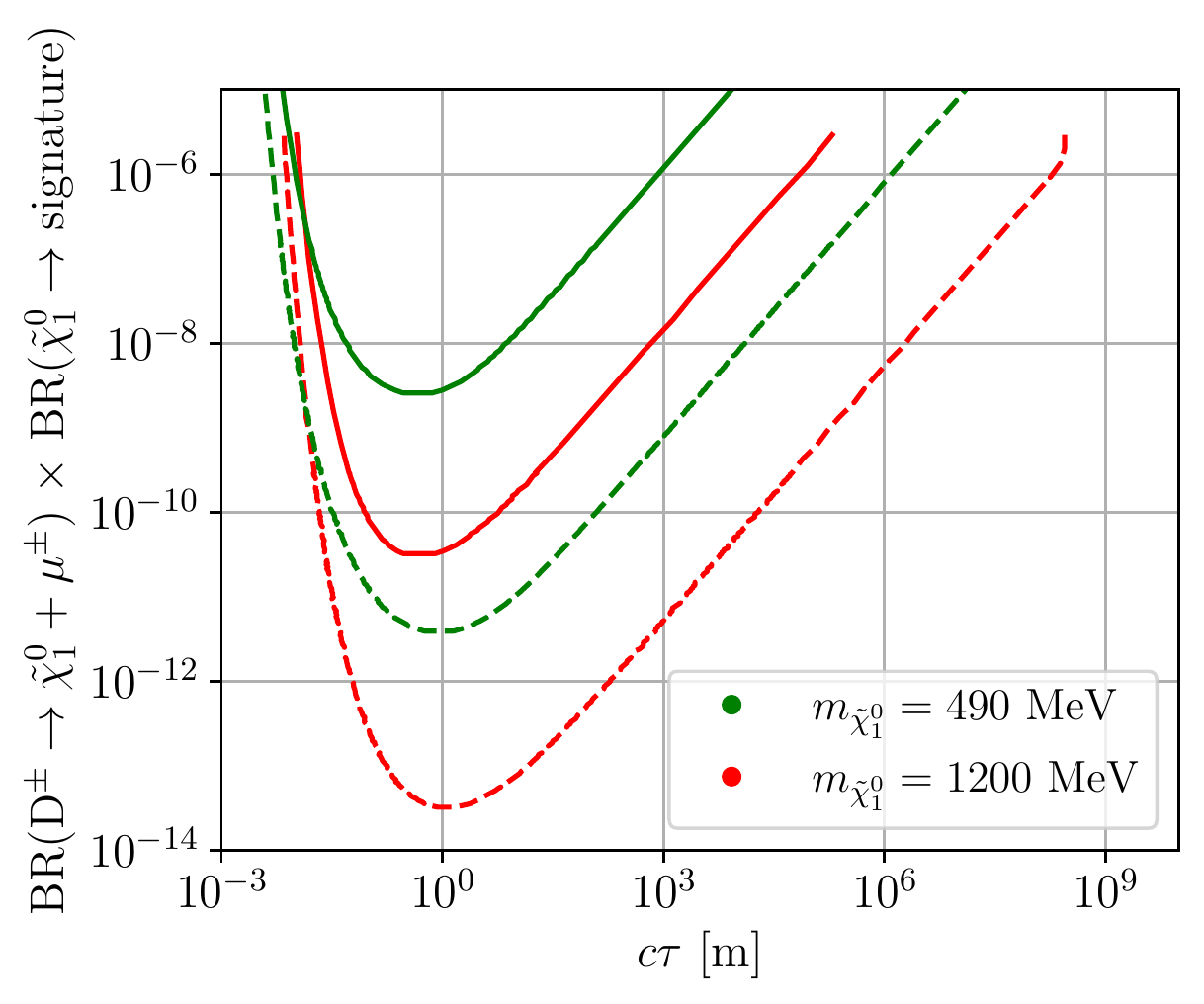}
	\caption{As in~\cref{fig::llp_neutralino::couplingvcoupling1} but for the benchmark scenario
		\textbf{B4} with $m_{\tilde{\chi}_1^0}=\SI{300}{\mega\electronvolt}$, \textit{cf}.~\cref{tab:benchmarks-rad-neutralino}.
		The right plot shows the sensitivity reach in BR$(D^\pm\to\tilde{\chi}^0_1+\mu^\pm)\times$BR$(\tilde{\chi}^0_1\to\text{signature})$ as a function of the neutralino decay length, $c\tau$, for $m_{\tilde\chi^0_1}=490$ and $\SI{1200}{\mega\electronvolt}$.}
	\label{fig::llp_neutralino::couplingvcoupling4}
\end{figure}

\begin{figure}[htb]
	\centering
	\includegraphics[width=0.495\textwidth]{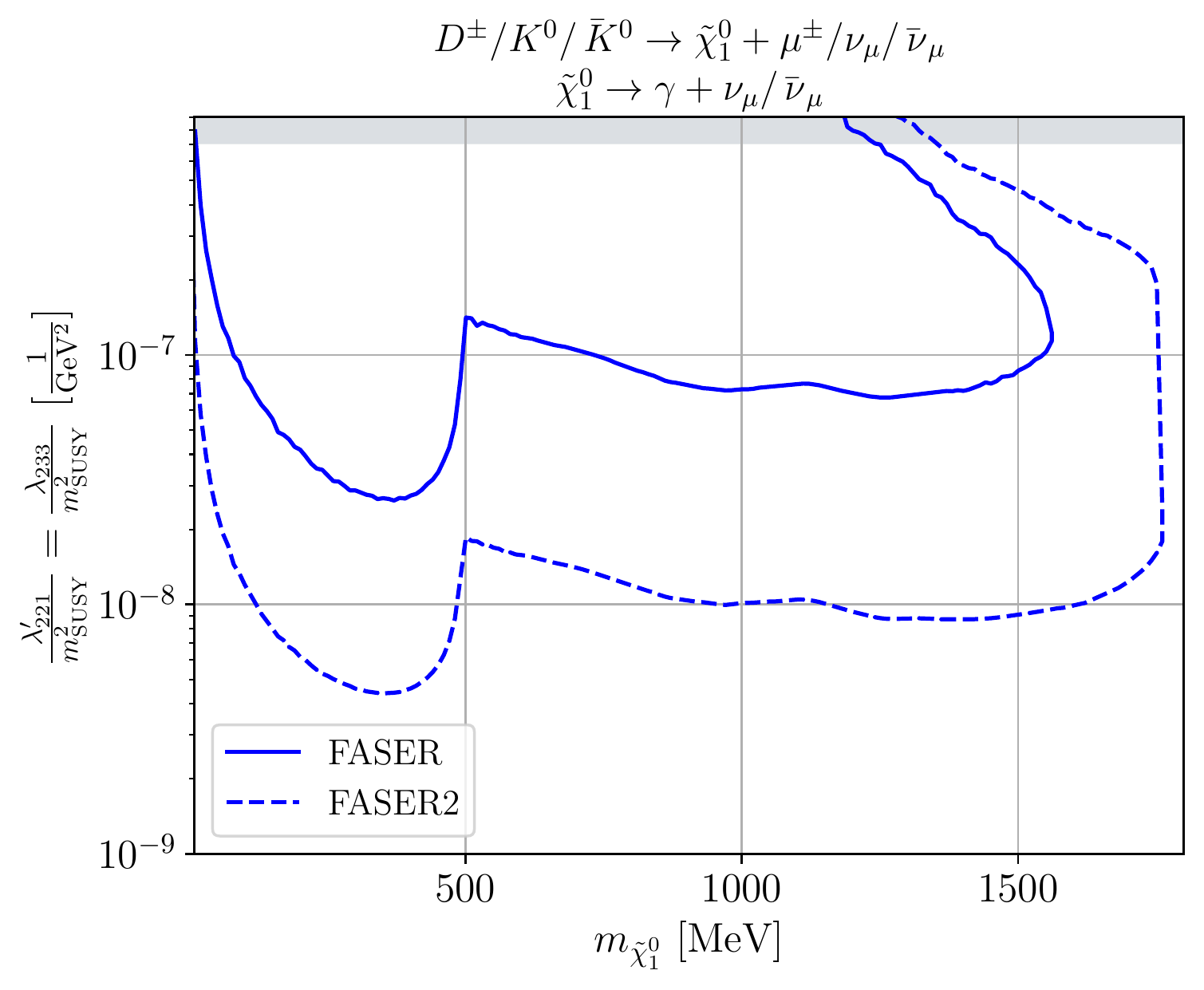}
	\caption{As in~\cref{fig::llp_neutralino::couplingvmass1} but for the benchmark scenario in \textbf{B4} while varying the neutralino mass. The sensitivity reach corresponds to \texttt{FASER} (solid line) and \texttt{FASER2} (dashed line).}
	\label{fig::llp_neutralino::couplingvmass4}
\end{figure}

In Figs.~\ref{fig::llp_neutralino::couplingvcoupling2} and
\ref{fig::llp_neutralino::couplingvmass2} (left), we display the 
sensitivity plots for benchmark scenario \textbf{B2}. The left plot 
of~\cref{fig::llp_neutralino::couplingvcoupling2} shows that now both 
\texttt{FASER} and \texttt{FASER2} have significant new reach in the 
couplings, $\lam^{\mathrm{P}}$ and $\lam^{\mathrm{D}}$, for $m_{\tilde{\chi}^0_1}=\SI{75}{\mega\electronvolt}$. The right plot in \cref{fig::llp_neutralino::couplingvcoupling2} looks similar 
to the right plot of~\cref{fig::llp_neutralino::couplingvcoupling1}, but it is now a plot of the
branching ratio product BR$(K^\pm\to\tilde\chi^0_1+\mu^\pm)\times$BR$(\tilde{\chi}^0_1\to\text{signature})$ versus the neutralino decay length, $c\tau$, and we have considered heavier LLP masses: $75$ and $\SI{300}{\mega\electronvolt}$.
Similar to \textbf{B1}, we overlap these results with existing bounds from searches for HNLs from kaon two-body decays, $K^\pm\to \mu^\pm + N$.
Refs.~\cite{Yamazaki:1984sj,NA62:2021bji} give the strongest current limits for HNL masses of \SI{75}{\mega\electronvolt} and \SI{300}{\mega\electronvolt}.
Ref.~\cite{Yamazaki:1984sj} is a peak search and bounds the HNL mixing-squared with the muon neutrino at $1.3\times10^{-5}$ for HNL mass of \SI{75}{\mega\electronvolt}.
Ref.~\cite{NA62:2021bji} searches for invisible particles and places a limit of $10^{-8}$ on the mixing-squared.
We convert these limits into bounds on BR$(K^\pm \to \mu^\pm + N)$ and obtain $10^{-5}$ and $2.4\times 10^{-8}$, respectively.
We depict these bounds in the right plot of~\cref{fig::llp_neutralino::couplingvcoupling2}, using BR$(\tilde{\chi}^0_1 \to \text{signature})=1$.
In particular, since Ref.~\cite{NA62:2021bji} is a missing-energy search, the limits are valid only for proper decay length larger than \SI{15}{\m}, as explicitly mentioned in the abstract of the paper.
We find that \texttt{FASER} and \texttt{FASER2} can probe the BR-product down to values significantly lower than these current limits.

In the neutralino mass-couping plane plot of~\cref{fig::llp_neutralino::couplingvmass2} (left), we observe
that the sensitivity at \texttt{FASER2} is reduced for lower masses compared to that in benchmark scenario \textbf{B1}, but, unlike the pion case, is robust over the entire higher-mass regime, right up to the kaon mass. This is because even though the charged decay production mode is kinematically forbidden beyond $m_{\tilde{\chi}^0_1}= m_{K^ {\pm}} 
-m_{\mu^{\pm}} \approx \SI{390}{\mega\electronvolt}$ -- leading to the small bump 
in the plot at that point -- the neutral decay mode has a comparable branching 
fraction. 

The two plots of $\lambda^\text{{P}}/m^2_{\text{SUSY}}$ vs.~$\lambda^{\text{D}}m^2_{\text{SUSY}}$ and branching ratio product vs.~$c\tau$ for \textbf{B3} with neutralino mass of $\SI{200}{\mega\electronvolt}$, are very similar to \cref{fig::llp_neutralino::couplingvcoupling2} for \textbf{B2} and are hence not shown explicitly here.
In \cref{fig::llp_neutralino::couplingvmass2} (right), we present the sensitivity plot for scenario \textbf{B3} for $\lam'_{112}=\lam^ 
{\text{P}} =\lam^{\text{D}}=\lam_{322}$, as a function of the neutralino mass. We
note that for $m_{\tilde{\chi}^0_1} \gtrsim 2m_{\mu}$, the decay mode 
$\tilde{\chi}^0_1\to \mu^\pm\mu^\mp\nu_{\tau} + \text{c.c.}$ opens up, leading to 
additional visible events. These are not included 
in~\cref{fig::llp_neutralino::couplingvmass2} (right).

\begin{figure}[h!]
	\centering
	\includegraphics[width=0.495\textwidth]{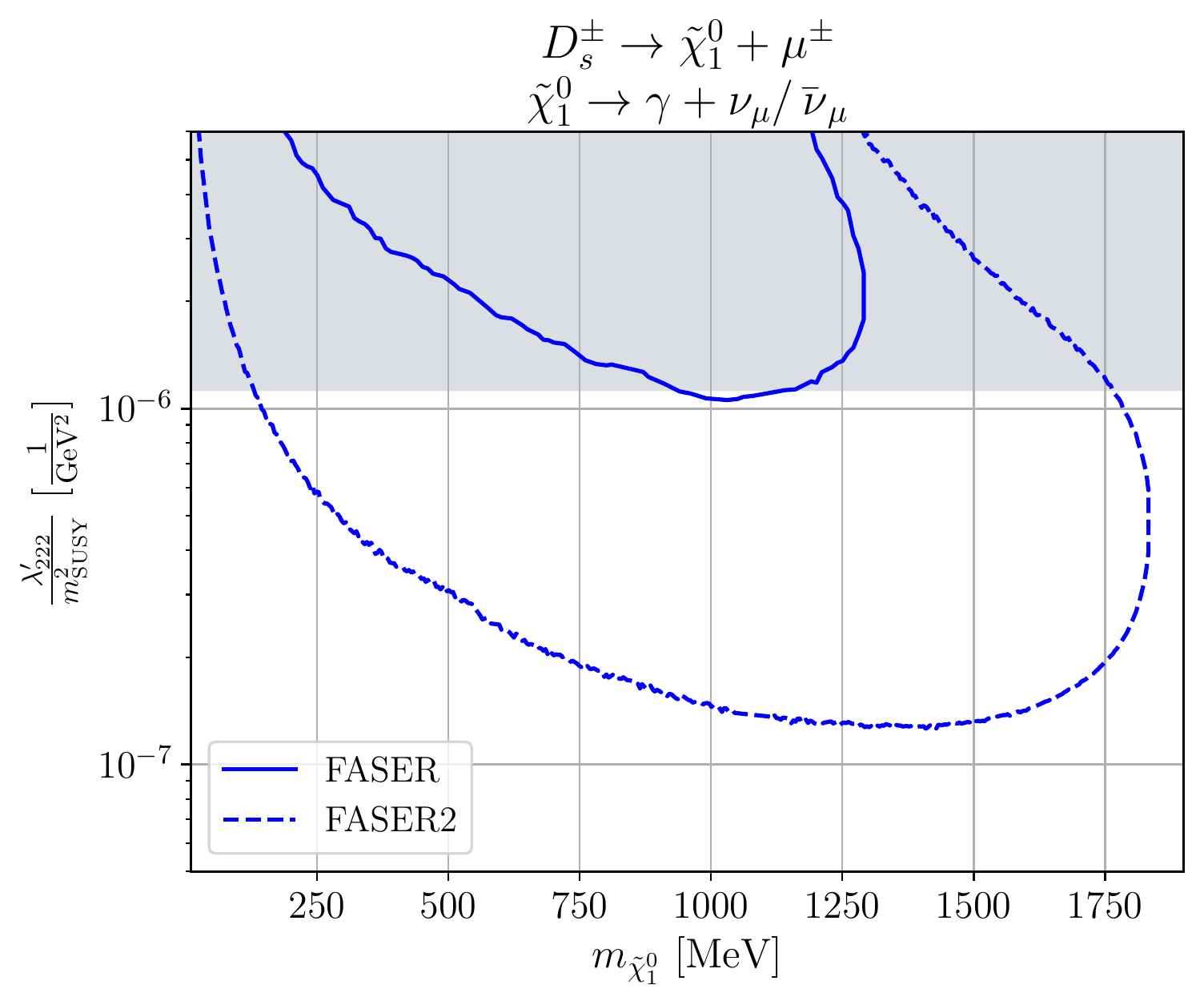}
	\includegraphics[width=0.495\textwidth]{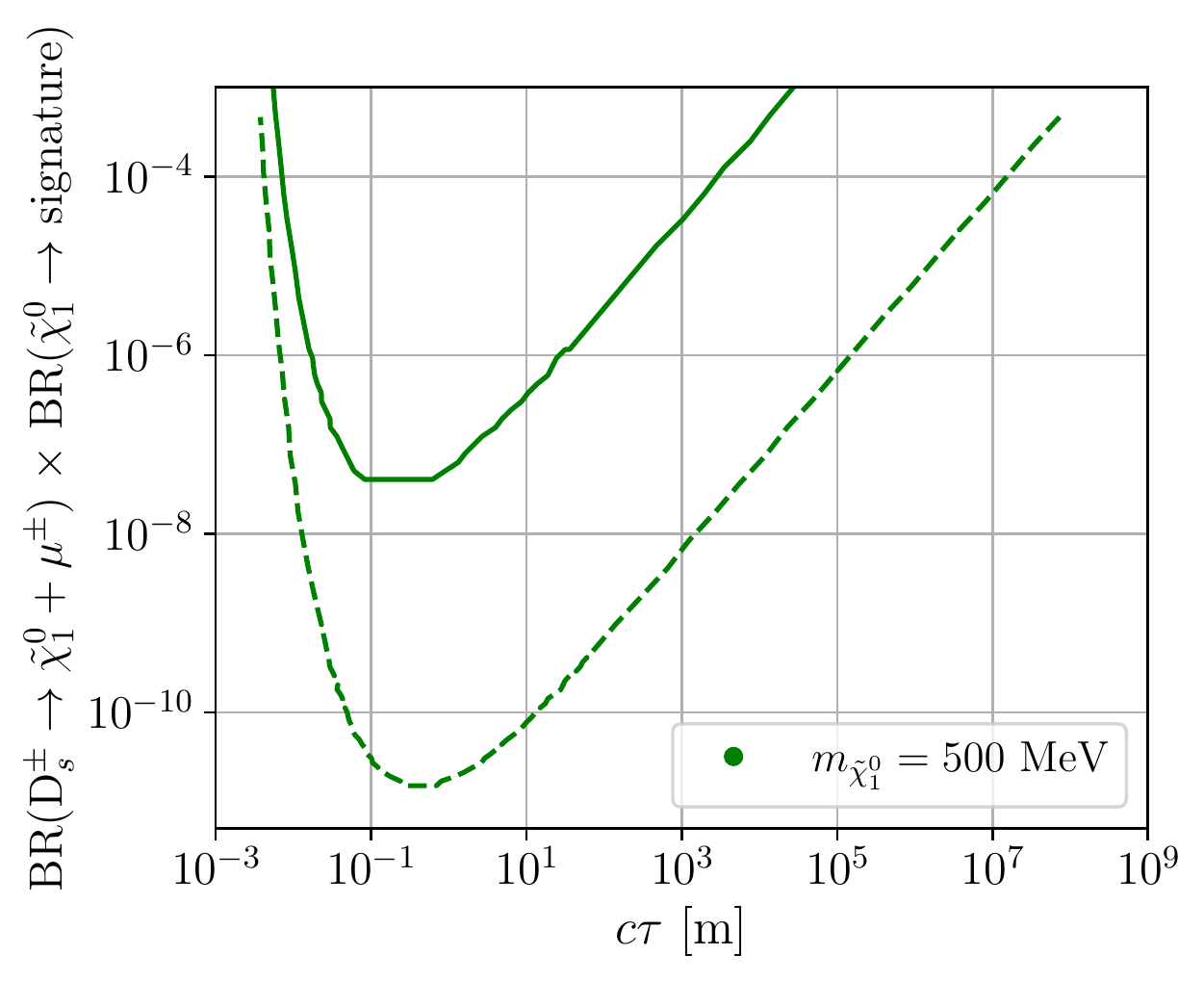}
	\caption{The left plot is as in~\cref{fig::llp_neutralino::couplingvmass1} but for the single coupling benchmark scenario in \textbf{B5} while varying the neutralino mass. The right plot shows the sensitivity reach 
		in BR$(D_S^\pm\to\tilde{\chi}^0_1+\mu^\pm)\times$BR$(\tilde{\chi}^0_1\to\text{signature})$ as a function of the neutralino decay length, $c\tau$, for $m_{\tilde\chi^0_1}=\SI{500}{\mega\electronvolt}$.
		The sensitivity reaches correspond to \texttt{FASER} (solid line) and \texttt{FASER2} (dashed line).}
	\label{fig::llp_neutralino::couplingvmass5}
\end{figure}

In Figs.~\ref{fig::llp_neutralino::couplingvcoupling4} and 
\ref{fig::llp_neutralino::couplingvmass4}, we show the corresponding
plots for the benchmark involving both $D$ and $K$ mesons -- scenario 
\textbf{B4}. One interesting feature in the mass-coupling plane is the
kink in the sensitivity curve near the kaon mass, $m_{K^0}\sim\SI{497}
{\mega\electronvolt}$. This is because for $m_{\tilde{\chi}^0_1}\gtrsim 
m_{K^0}$, the kaon production mode, $K^0/\bar{K}^0 \to \tilde{\chi}^0_1 
+ \nu_{\mu}/\bar{\nu}_{\mu}$ switches off. Since kaons are more
abundant than $D$ mesons at the LHC, this leads to reduced senstivity 
beyond this threshold. For larger masses, the neutralino has decay modes 
into $K^0$ and $K^{*0}$ plus neutrino opening up at the respective mass 
thresholds; as before, we only count the photon events as signal. 
There is an additional interesting feature for this scenario: The 
sensitivity curve starts to `turn back' in the large coupling, large mass
region indicating a drop in sensitivity. This happens as the lifetime of 
the neutralino becomes too short, decaying well before reaching \texttt{FASER} or \texttt{FASER2}; this effect is made more acute by the additional decay modes that open up.
To our knowledge, there are no existing searches for HNLs in $D^\pm\to \mu^\pm+N$ decays; therefore, we do not place any existing bounds in the right plot of Fig.~\ref{fig::llp_neutralino::couplingvcoupling4}.

\cref{fig::llp_neutralino::couplingvmass5} shows the sensitivity reach 
for scenario \textbf{B5}, where only one RPV coupling is switched on (thus, there is no coupling-coupling plane plot).
The left plot shows the sensitivity reach of \texttt{FASER} and \texttt{FASER2} in the mass-coupling plane.
Once again, for the plots, we do not consider the additional decay modes into $\eta, \eta'$ or $\phi$ plus neutrino that open up at the respective mass thresholds, for our signature.
The large mass, large coupling regime has reduced sensitivity for the same reason stated above.
The right plot then contains the sensitivities of \texttt{FASER} and \texttt{FASER2} to the decay branching fraction product as a function of $c\tau$ for $m_{\tilde{\chi}^0_1}=\SI{500}{\mega\electronvolt}$.
For this plot, as in \textbf{B4}, there is no existing limit that can be obtained from an HNL search in $D^\pm_s\to \mu^\pm + N$ decays.

\begin{figure}[h!]
	\centering
	\includegraphics[width=0.495\textwidth]{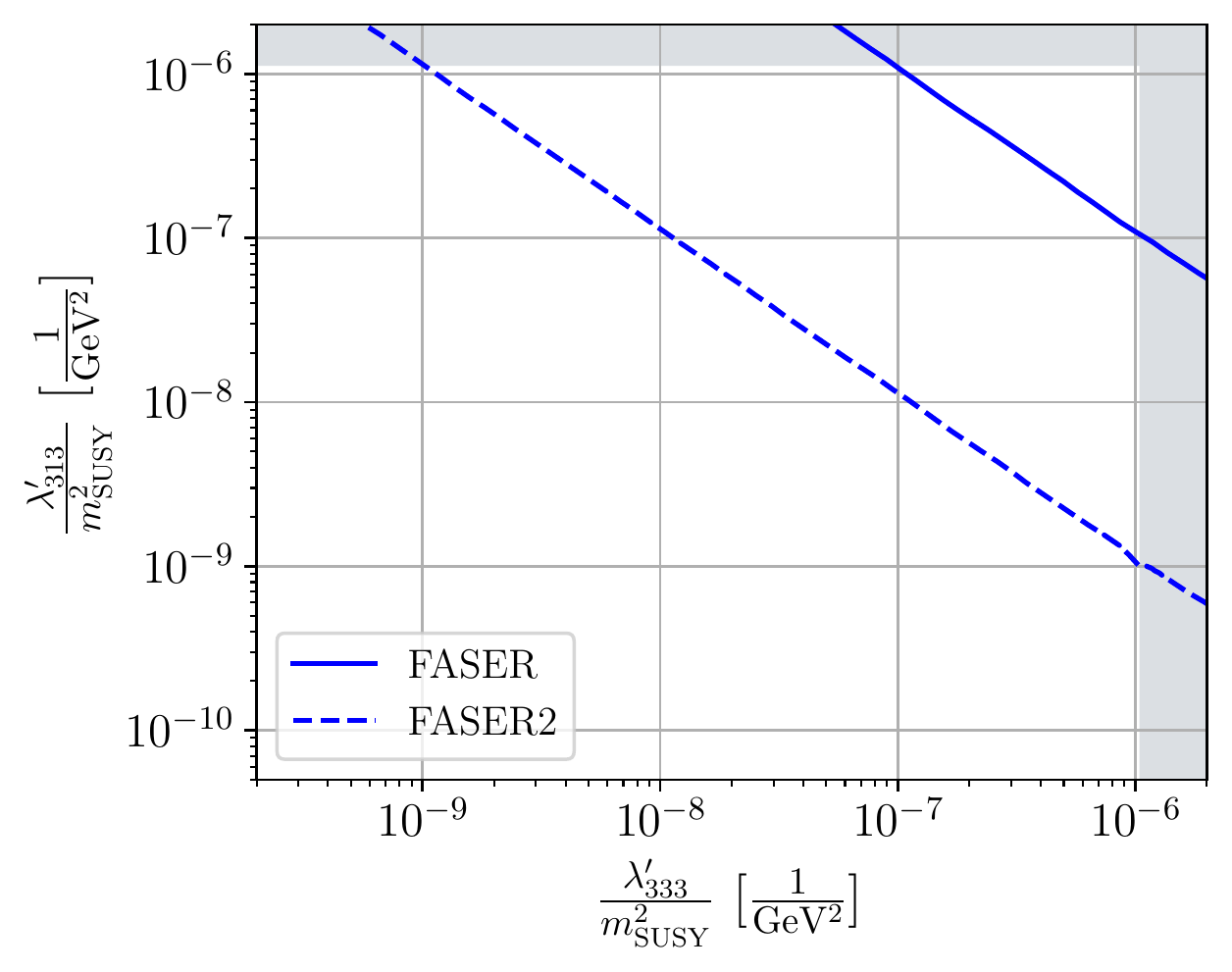}
	\includegraphics[width=0.495\textwidth]{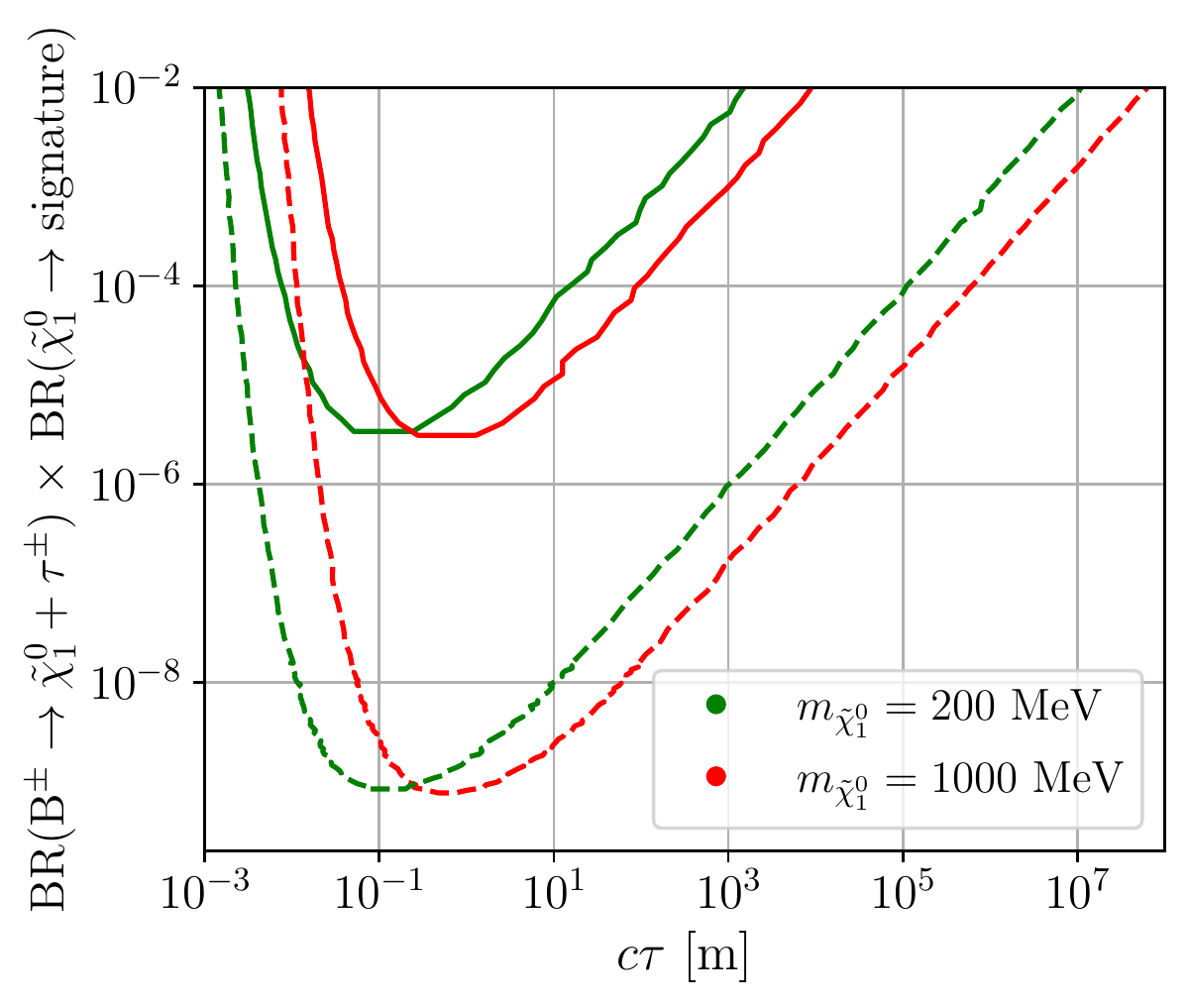}
	\caption{As in~\cref{fig::llp_neutralino::couplingvcoupling1} but for the benchmark scenario \textbf{B6} with $m_{\tilde{\chi}_1^0}=\SI{1}{\giga\electronvolt}$, \textit{cf.}~\cref{tab:benchmarks-rad-neutralino}. The right plot shows the sensitivity reach in BR$(B^\pm\to\tilde
		{\chi}^0_1+\tau^\pm)\times$BR$(\tilde{\chi}^0_1\to\text{signature})$ as a function of the neutralino decay length, $c\tau$,
		for $m_{\tilde\chi^0_1}=200$ and $\SI{1000}{\mega\electronvolt}$.}
	\label{fig::llp_neutralino::couplingvcoupling6}
\end{figure}

\begin{figure}[htb]
	\centering
	\includegraphics[width=0.495\textwidth]{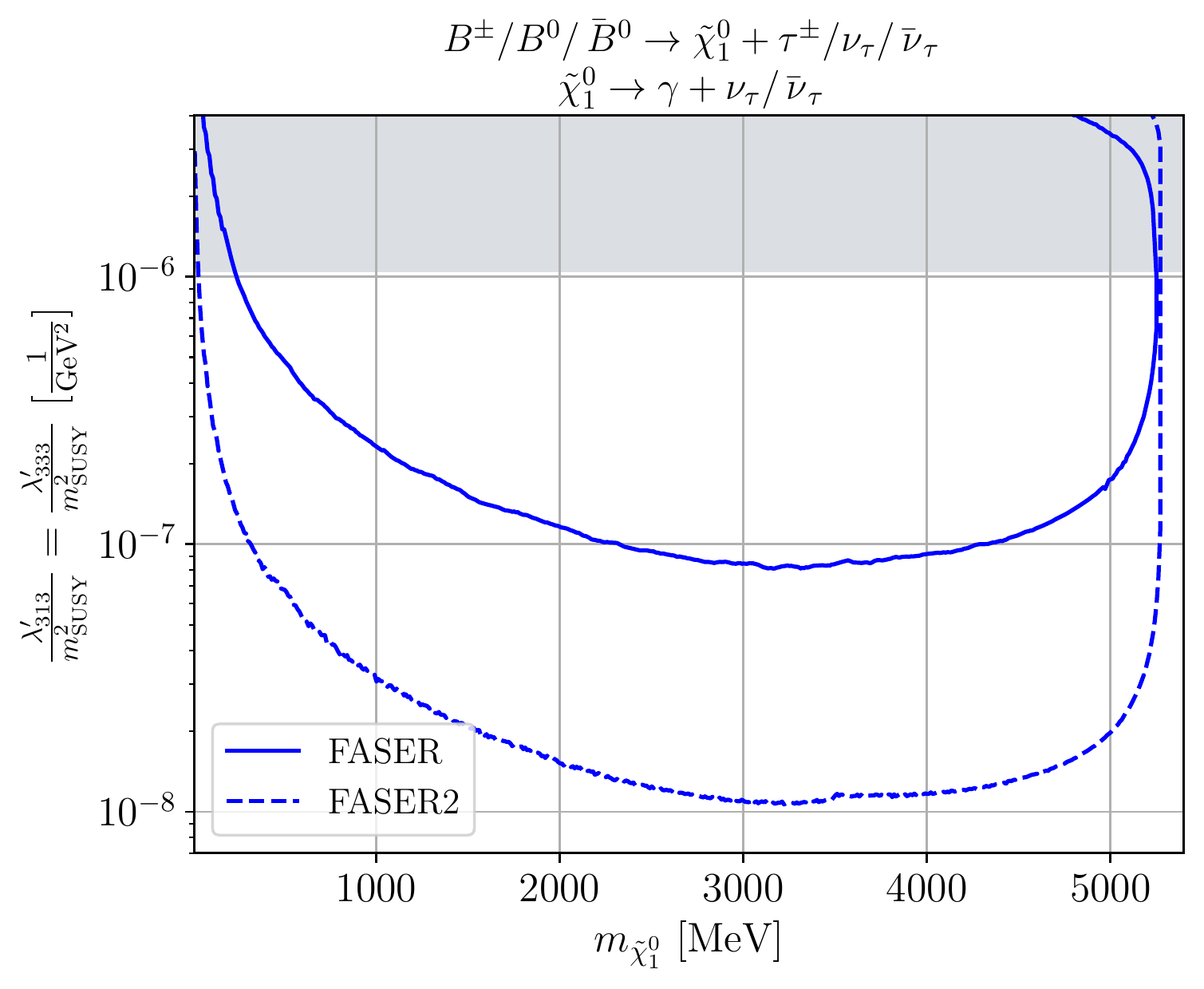}
	\caption{As in~\cref{fig::llp_neutralino::couplingvmass1} but for the benchmark scenario in \textbf{B6} while varying the neutralino mass. The sensitivity reach corresponds to \texttt{FASER} (solid line) and \texttt{FASER2} (dashed line).}
	\label{fig::llp_neutralino::couplingvmass6}
\end{figure}

Finally, Figs.~\ref{fig::llp_neutralino::couplingvcoupling6} and \ref{fig::llp_neutralino::couplingvmass6} contain the sensitivity plots for benchmark scenario \textbf{B6}, involving $B$ mesons.
We observe in the right plot of Fig.~\ref{fig::llp_neutralino::couplingvcoupling6}, that the reaches in the branching ratio product are weaker than those in the previous scenarios, because the production rates of the $B$ mesons are orders-of-magnitude smaller than those of the lighter mesons at the LHC.
As in the previous two benchmark scenarios, we do not find an existing search for HNLs in $B^\pm\to \tau^\pm+N$ decays that could be recast into bounds relevant to us, in the BR-product vs.~$c\tau$ plane.
In Fig.~\ref{fig::llp_neutralino::couplingvmass6}, as before, we see that the sensitivity in the 
mass-coupling plane is robust across the kinematically allowed mass range
since the neutral mode is available even when the charged mode is 
switched off for $m_{\tilde{\chi}^0_1} \geq m_{K^{\pm}}- m_{\tau
	^{\pm}}$. This time the drop in sensitivity in the large mass, large 
coupling region is milder compared to the previous two cases as there are
no additional decay modes contributing.

\section{Conclusions}\label{sec:conclusions}

We have estimated the sensitivity reach of \texttt{FASER} and \texttt{FASER2} at the LHC, for a sub-GeV bino-like lightest
neutralino decaying to a photon in the context of R-parity-violating supersymmetry.
With R-parity broken, the lightest neutralino can be lighter than the GeV scale, or even massless, without 
violating observational and experimental bounds, as long as it decays.
Assuming the lightest neutralino is the lightest supersymmetric particle (LSP), it can be produced 
from rare meson decays, and can decay to a photon and a neutrino, via certain RPV couplings. 

In the sub-GeV mass range, since the RPV couplings are required to be small by existing constraints, the bino-like neutralino is expected to have a long lifetime. Once produced from mesons' decays at the LHC, it is highly boosted in the very forward direction.
Therefore, we have chosen to focus on the experimental setups of \texttt{FASER} and \texttt{FASER2} for observing the single-photon signature resulting inside the detector decay volumes.

We have considered several theoretical benchmark scenarios and performed Monte 
Carlo simulations in order to determine the projected sensitivity reaches at 
\texttt{FASER} and \texttt{FASER2}. Our study has found that these experiments are sensitive to parameter space beyond the current bounds by orders of magnitude.

\bigskip
\section*{Acknowledgments}
\bigskip
We thank Florian Domingo, Max Fieg, Oliver Fischer, Julian G\"unther, Martin Hirsch, Felix Kling, and Guanghui
Zhou for useful discussions. Z.S.W. is supported by the Ministry of 
Science and Technology (MoST) of Taiwan with grant number 
MoST-110-2811-M-007-542-MY3. HKD, DK, and SN acknowledge partial 
ﬁnancial support by the Deutsche Forschungsgemeinschaft (DFG, German 
Research Foundation) through the funds provided to the Sino-German 
Collaborative Research Center TRR110 “Symmetries and the Emergence of 
Structure in QCD” (DFG Project ID 196253076 - TRR 110).

\bigskip

\bibliographystyle{JHEP}
\bibliography{bibliography}
\end{document}